\documentclass[12pt]{iopart}

\usepackage{cite}
\usepackage{graphicx}

\usepackage{color}
\usepackage{ulem}

\newcommand{\be}{\begin{eqnarray}}
\newcommand{\ee}{\end{eqnarray}}
\newcommand{\sub}[1]{_\mathrm{#1}}
\newcommand{\s}{\sigma}

\begin{document}

\title[Dynamic force spectroscopy of DNA hairpins. II.]{Dynamic force spectroscopy of DNA hairpins. II. Irreversibility and dissipation}

\author{M~Manosas$^1$\footnote{Present address: Laboratoire de Physique Statistique, Ecole Normale Sup\'erieure, Unit\'e Mixte de Recherche 8550 associ\'ee au Centre National de la recherche Scientifique et aux Universit\'es Paris VI et VII, 24 Rue Lhomond, 75231 Paris, France.}, A~Mossa$^1$,  N~Forns$^{1,2}$, J~M~Huguet$^1$ and F~Ritort$^{1,2}$}

\address{$^1$ Departament de F\'{\i}sica Fonamental, Facultat de F\'{\i}sica, Universitat de Barcelona, Diagonal 647, 08028 Barcelona, Spain}
\address{$^2$ CIBER-BBN Networking center on Bioengineering, Biomaterials and Nanomedicine}
\ead{ritort@ffn.ub.es}

\begin{abstract}
We investigate irreversibility and dissipation in single
molecules that cooperatively fold/unfold in a two state manner under the action of
mechanical force. We apply path thermodynamics to derive analytical
expressions for the average dissipated work and the average hopping
number in two state systems. It is shown how these quantities only depend on two
parameters that characterize the folding/unfolding kinetics of the molecule:
the fragility and the coexistence hopping rate. The latter has to be
rescaled to take into account the appropriate experimental
setup. Finally we carry out pulling experiments 
with optical tweezers in a specifically designed DNA hairpin that
shows two-state cooperative folding. We then use these experimental results to validate 
our theoretical predictions.  
\end{abstract}

\pacs{82.37.Rs, 87.80.Nj}


\maketitle

\section{Introduction}

One of the most exciting aspects of single molecule techniques is the
possibility of accurately measuring tiny amounts of
energy down to a tenth of a $k\sub{B}T$ (at room temperature $1\ k\sub{B}T$
corresponds to approximately 4 pN$\cdot$nm which is on the order of $10^{-21}\ 
J$) \cite{Ritort06}. The possibility of measuring such 
small energies opens new perspectives in the exploration of the
properties and behavior of biological matter at both cellular and
molecular levels.   Questions such as
irreversibility, dissipation and energy fluctuations have received a
new boost under the heading of {\it nonequilibrium thermodynamics of
  small systems}, a discipline that succesfully combines theory and
experiments and uses individual biomolecules as models to investigate
the energetics of complex systems \cite{BusLipRit05,Ritort08}.

\begin{figure}
\begin{center}
\includegraphics[width=10cm,angle=90]{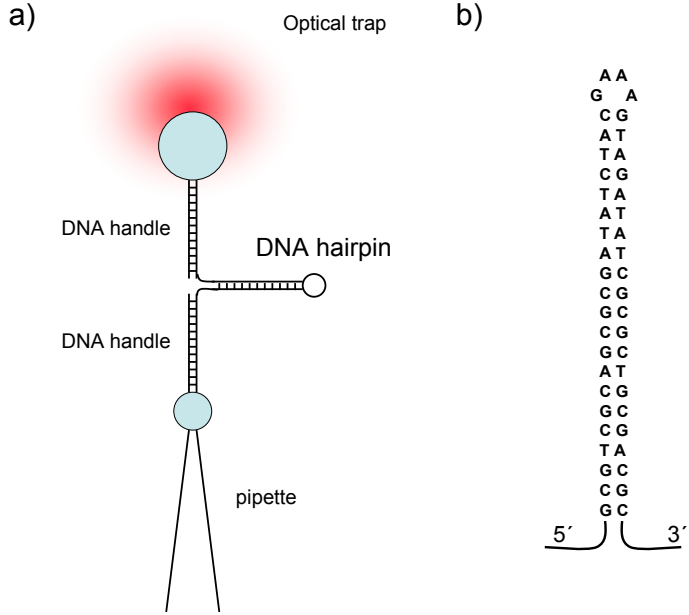}
\caption{ (a) Experimental set up. (b) DNA sequence.} 
\label{fig1}
\end{center}
\end{figure}

This paper is a continuation of a companion one where we
investigated in detail questions related to thermodynamics and
kinetics of simple
two-state hairpins under applied force \cite{MosManForHugRit08}.
Here we investigate issues related to irreversibility and
dissipation in pulling experiments of DNA hairpins that fold/unfold
under the action of mechanical force. In these experiments the molecule is
repeatedly unfolded/folded by increasing/decreasing the force at a
given pulling rate in a controlled way.  The output of such
experiments is the force-distance curve (FDC), a diagram that shows
the force as a function of the trap position. We have carried out
experiments (experimental setup shown in \fref{fig1}(a)) using a high stability newly designed miniaturized
dual-beam optical tweezers apparatus \cite{Smith08}
to pull a specifically designed DNA hairpin sequence (\fref{fig1}(b)). The specific sequence shows a
simple two-state free energy landscape, ideal to compare theory and
experiments \cite{MosManForHugRit08}. We use the recently introduced
theoretical approach of path thermodynamics \cite{Rit22,Ritort08} to investigate
questions related to the irreversibility and energy dissipation in two state systems. The study of DNA
hairpins \cite{WooBehLarTraHer06} has advantages compared to RNA
studies because the former degrades much slower than RNA does. This makes
DNA an excellent model to address physics related questions. 

The content of the paper is divided into two main
parts: the first part is mainly theoretical and describes the analytical
results; the second part presents the experimental results and compares
them with the theoretical predictions. The paper ends with some
conclusions followed by a few appendices that include some technical
aspects of the analytical computations.

\section{Mechanical work, hopping number and free energy landscapes: a short reminder}
\label{2-s}
\begin{figure}
\begin{center}
\vspace{0.9cm}
\includegraphics[width=10cm,angle=0]{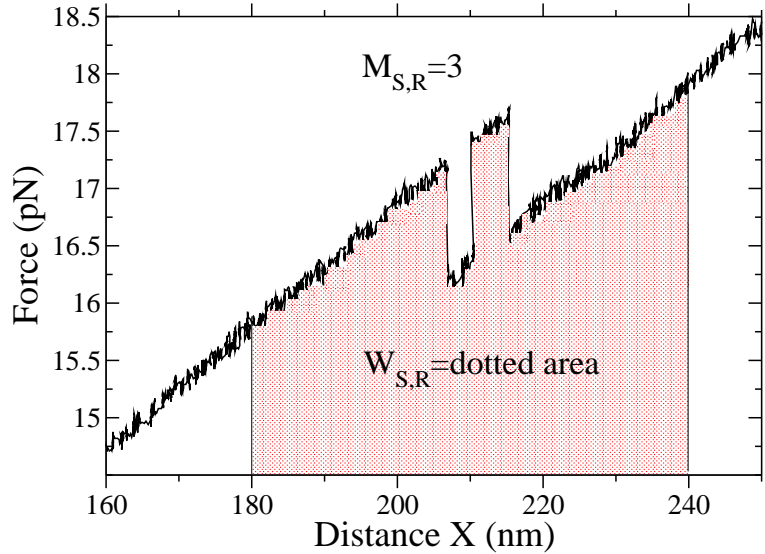}
\caption{ A representative FDC corresponding to either a stretching or releasing
part of a cycle. It shows the relevant quantitites that we investigate: the mechanical work exerted on the molecule $W\sub{S(R)}$ (shown
as the dotted area); and the number of transitions the molecule executes
between the two states, $M\sub{S(R)}$ ( $=3$ in the example of the figure).} 
\label{fig2}
\end{center}
\end{figure}

We consider a pulling cycle ($\Gamma$) consisting of a stretching (S)
and a releasing (R) parts. In the stretching part the force is
increased from a low force $f_{\rm min}$ where the hairpin is always
folded to a high force $f_{\rm max}$ where the hairpin is always
unfolded. In the releasing part the force is decreased from $f_{\rm
  max}$ back to $f_{\rm min}$. To quantify
irreversibility and energy dissipation during the stretching and
releasing parts of the cycle, we introduce the two following
quantities:

\begin{itemize}

\item The mechanical work $W\sub{S(R)}$. The mechanical work $W\sub{S(R)}$ along the
stretching (releasing) part of the cycle $\Gamma$ is the area below the
FDC between the positions $X_{\rm min}$ and $X_{\rm max}$
(corresponding to the previously defined forces $f_{\rm min}$ and $f_{\rm max}$ along the
FDC). It is defined by
\be
W\sub{S(R)}(\Gamma)=\int_{X_{\rm min}}^{X_{\rm max}}\rmd X F\sub{S(R)}(X) \,,
\label{intro1}
\ee
where the subindex S (R) refers to the stretching (releasing) part
of the cycle.  Throughout this work we will take $W\sub{S}$ and $W\sub{R}$ as
positive and negative quantities respectively although, according to
\eref{intro1}, both have positive signs. In fact, the work in the S part
of the cycle is performed by the instrument on the system ($\rmd X>0$)
whereas in the other case (R) the work is returned by the system to
the instrument ($\rmd X<0$). \Fref{fig2} shows how we measure the
  work along a given FDC. 

\item The hopping number $M\sub{S(R)}$. The hopping number $M\sub{S(R)}$ along the
stretching (releasing) part of the cycle $\Gamma$ is the number of
transitions or jumps (corresponding to force rips) that the molecule executes
during the stretching (releasing) part of the cycle. $M\sub{S(R)}$
can take only odd values ($1,3,5,\dots$) for all paths $\Gamma$ performed
between $f_{\rm min}$ and $f_{\rm max}$. An illustration is shown in \fref{fig2}.

\end{itemize}

These are the main quantities that we will focus on in this paper. Out
of the total work $W^{\rm S(R)}$ we can also extract, for each cycle
$\Gamma$, the dissipated work $W_{\rm dis}^{\rm S(R)}$ which is equal
to the difference between the work and the reversible work. The
reversible work can be estimated either using the Jarzynski equality
along the S (R) process \cite{Jarzynski97} or the Crooks fluctuation
relation \cite{Crooks99,ColRitJarSmiTinBus05}. The latter combines
measurements from both the S and R processes and provides better
estimates for free energy differences. The Crooks fluctuation relation
has been applied to recover the free energy of the DNA sequence shown
in \fref{fig1}(b) (see our companion paper \cite{MosManForHugRit08}).

\begin{figure} 
\begin{center}
\includegraphics[scale=0.67,angle=0]{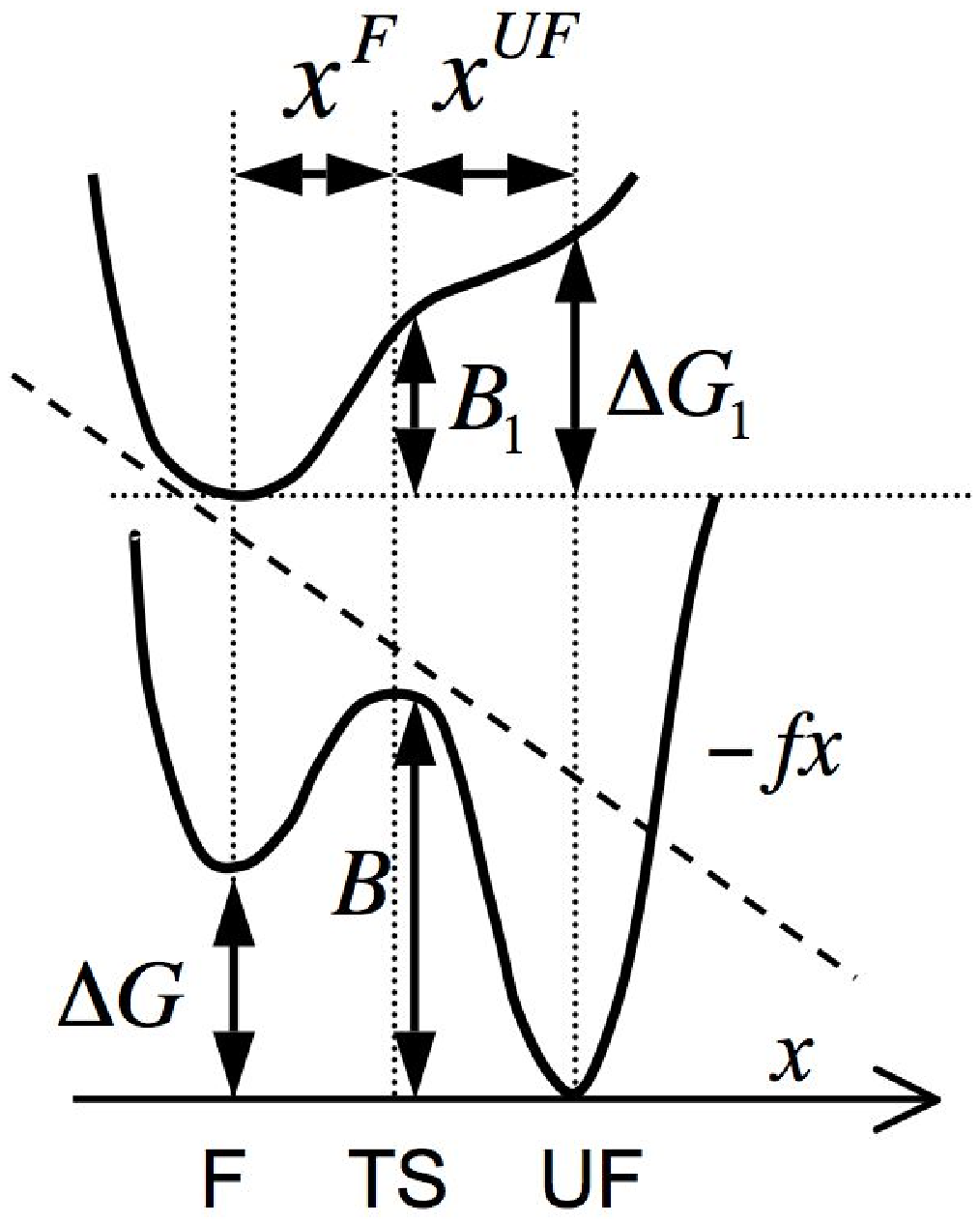} \includegraphics[scale=0.37,angle=0]{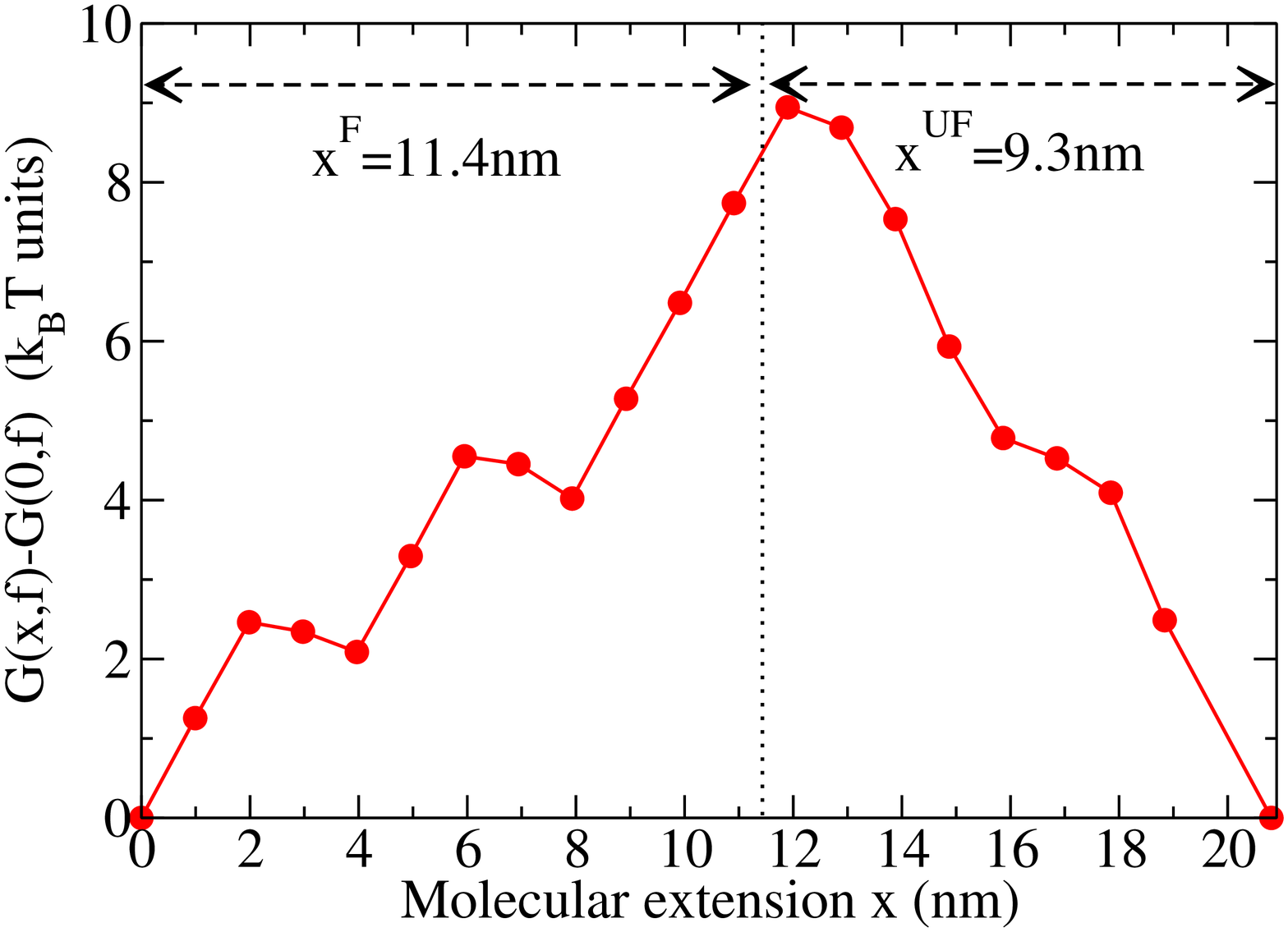} 
\caption{ (Upper panel) Schematic picture of the two-state model. The free energy
landscape of the molecule along the reaction coordinate axis $x$ has two minima corresponding to
the F and UF states. These states have free
energies $G\sub{F}$ and $G\sub{UF}$ with $\Delta G=G\sub{UF}-G_{\rm
  F}$. The two states (F and UF) are separated by the transition state
(TS) that has a free energy $B$ higher than that of the F state. The
TS is located at distances $x^{\rm{F}}$ and $x^{\rm{UF}}$ from the F
and UF states respectively. When a mechanical force is applied to the
ends of the molecule the free energy landscape is tilted along $x$,
decreasing the free energy of the UF state and the TS relative to the
F state. This induces a variation of $B$ and $\Delta G$ which, in the
simplified Kramers--Bell picture are given by $B=B_1-fx^{\rm F};\Delta
G=\Delta G_1-fx\sub{m}$ where $B_1,\Delta G_1$ are constant free energy
parameters.  (Lower panel) Free energy landscape $G(x,f)$ for the
DNA hairpin shown in \fref{fig1}(b) plotted as a function of the
number of open base pairs at $f=17.9$ pN (buffer conditions are 1M NaCl
and $23^{\circ}$ C). The landscape has been calculated using the
nearest neighbour model \cite{SantaLucia1998} for thermodynamics using the free energy
parameters given by Mfold \cite{Zuker2003} and the theory developed in
\cite{MosManForHugRit08}.}
\label{fig3} 
\end{center} 
\end{figure}

The mechanical folding and unfolding of small nucleic acid (DNA or
RNA) hairpins is commonly described with a two-state model
\cite{Bon,Fern,Hum1,pan1,Lip1}. In this model the hairpin can adopt two
conformation or states, the folded (hereafter referred as F) and the
unfolded or stretched (hereafter referred as UF) state.  When subject
to force, the projection $x$ of the molecular extension along the
force axis is an adequate reaction coordinate for the
folding-unfolding reaction. For a given applied force $f$, the
two-state approach considers a single-kinetic pathway for the
unfolding/folding reactions and the free energy landscape is
characterized by a single transition state (hereafter referred as TS).
The TS is the state with highest free energy along the reaction
coordinate and determines the kinetics of the unfolding (folding)
reaction. The simplest version of the Kramers--Bell model (see
\cite{MosManForHugRit08} for details) is schematically depicted in
 \fref{fig3}(a). It involves only four parameters: the free energy
difference $\Delta G_1$ the kinetic barrier $B_1$ and the distances
$x^{\rm{F}}$ and $x^{\rm{UF}}$ along the reaction coordinate axis that
separates the TS from the F and UF states respectively. The total
distance between the F and U states will be denoted as $x\sub{m}$ and can
be written as $x\sub{m}=x^{\rm F}+x^{\rm UF}$.

Two-state folders are well described by the so-called simplified
Kramers--Bell kinetic rates,
\be
k_{\rightarrow}(f)=k\sub{m}\exp{\left(\beta f x^{\rm F}\right)}\,, \qquad k_{\leftarrow}(f)=k\sub{m}\exp{\left(\beta(\Delta G_1-fx^{\rm UF})\right)}\,,
\label{e7}
\ee
with $\beta=1/k\sub{B}T$, $T$ being the temperature of the environment and
$k\sub{B}$ the Boltzmann constant. $k\sub{m}$ corresponds to the unfolding
rate at zero force and is given by $k\sub{m}=k_0\exp{-(\beta B_1)}$ where
$k_0$ is an attempt frequency describing microscopic oscillations and
effects of the experimental setup (bead motion, fluctuations in
extension of handles and ssDNA), and $B_1$ is the kinetic barrier. The
parameter $\Delta G_1$ in \eref{e7} is related to the total free energy
difference between the F and UF states at force $f$: $\Delta
G(f)=\Delta G_1-fx\sub{m}$. $\Delta G_1$ gets contributions from the free
energy of formation of the hairpin at zero force, $\Delta G_0$, and
the free energy of stretching the ssDNA up to the
total extension $x\sub{m}$ ($=x^{\rm F}+x^{\rm UF}$) at force $f$.

 \Eref{e7} defines the coexistence force $f\sub{c}$ where both rates
 are equal, $k_{\rightarrow}(f\sub{c})=k_{\leftarrow}(f\sub{c})$, giving
 $f\sub{c}=\Delta G_1/x\sub{m}$. All parameters entering into the simplified rates
 \eref{e7} (i.e. $k\sub{m}$, $x^{\rm F}$, $x^{\rm UF}$, $\Delta G$) are taken
 to be independent of force and
 $\Delta G_1$ is evaluated at the critical force
 $f\sub{c}$ (see \cite{MosManForHugRit08} for a thorough discussion).

The distances $x^{\rm F},x^{\rm UF}$ characterize how much the molecule deforms before unfolding. It is common to define the fragility parameter $\mu$ \cite{Lef53,HyeThir05,HyeThir06,man1}: 
\be
\mu=\frac{x^{\rm F}-x^{\rm UF}}{x^{\rm F}+x^{\rm UF}}\,.
\label{mu}
\ee
$\mu$ lies in the range $[-1:1]$ and defines the degree of compliance of
the molecule under the effect of tension. Fragile or compliant molecules are those in which
$x^{\rm F}$ is larger than $x^{\rm UF}$ and $\mu$ is positive. In
contrast, when $x^{\rm UF}$ is larger than $x^{\rm F}$ and $\mu$ is
negative we speak of brittle structures. 

\section{Mechanical work and hopping number in path thermodynamics}
\label{Ana} 

In this section we apply path thermodynamics \cite{Ritort08} to find analytical expressions
for the average dissipated work and the average hopping number in
two-state systems. These are later compared
with our experimental results obtained in DNA hairpins. The general
reader not interested in the mathematical details of the derivation can skip \sref{derivation}.

\subsection{Derivation of the equations.}
\label{derivation}
Let us consider a schematic version of our two-state system represented
by a discrete spin variable $\sigma$ that labels the state of the system,
$\sigma=-1,1$. The spin $\sigma$ can switch between states of 
zero free energy ($\sigma=-1$) and free energy $\Delta G$ ($\sigma=1$). The Hamiltonian of the system can be written as:
\be
H(\sigma) =\Delta G \frac{(\sigma +1)}{2} \,. 
\label{e1}
\ee
In the presence of an external field $f$ the Hamiltonian has a new term and is
 given by:
\be
H(\sigma) =\Delta G\frac{(\sigma +1)}{2}-x\sub{m}f\frac{(\sigma +1)}{2} \,, 
\label{e2}
\ee
where $x\sub{m}$ is the coupling between the system and the external
field.  For the case of the folding-unfolding reaction under force $f$, the two
states, $\sigma=-1,1$, correspond to the folded (F) and the unfolded
(UF) states respectively, and $x\sub{m}$ is the molecular distance
separating them along the reaction coordinate axis,
i.e. $x\sub{m}=x^{\rm{F}}+x^{\rm{UF}}$ (see \fref{fig3}).

We consider an isothermal perturbation where the external field is
changed from an initial value $f\sub{min}$ to a final one $f\sub{max}$
following a given protocol $f(t)$, denoted as a ramping protocol. In
this protocol the force is externally controlled and does not
fluctuate. The equilibrium and thermodynamic properties that are
computed when the force is controlled correspond to the so-called
force ensemble.

According to the formalism of path thermodynamics introduced in
\cite{Rit22}, we consider a system composed of $N$ independent two-state
particles with dynamical evolution governed by the Hamiltonian in
\eref{e2}.  A path or trajectory $\Gamma$ of the system is defined by
the sequence of configurations $C_{k}$, $\Gamma=\{ C_{k};1\leq k \leq
N\sub{s}\}$ with $C_{k}=\lbrace\sigma_{i}^{k},1\leq i \leq N\rbrace$, the index $k$ denoting a time equal to $k\Delta t$ where the total time
$t_{\rm total}$ has been discretized in $N\sub{s}$ steps of duration $\Delta
t=t_{\rm total}/N\sub{s}$.  The work $W$ per particle along a single path
$\Gamma$ reads:
\be
W(\Gamma )=-\frac{x\sub{m}}{2}\sum_{k=0}^{N\sub{s}-1}m^{k+1}(f^{k+1}-f^{k})\,, \label{Ba} 
\ee 
where $m^{k}=\frac{1}{N}\sum_{i=1}^{N}\sigma_{i}^{k}$ and $f^{k}$ are
the values of the magnetization and external field at time $k$.  The
factor 1/2 in \eref{Ba} arises from the absolute value of the change in extension associated to
the transition $\s\to-\s$, $\Delta(x\sub{m}\sigma/2)=x\sub{m}$. On the
other hand, the hopping number $M$ per particle over a single path
reads: 
\be 
M(\Gamma)=\frac{1}{N}\sum_{k=0}^{N\sub{s}-1}\sum_{i=0}^{N}\frac{1}{2}(1-\sigma_{i}^{k}\sigma_{i}^{k+1})
\,.
\label{Bb} 
\ee 
The distribution of probability $P_{N}(\theta)$ for any observable
$\theta$ (e.g. $W$ or $M$) measured over all  paths can
be written as: 
\be 
P_{N}(\theta )=\sum_{\{ \Gamma \} }\delta (\theta -
\theta (\Gamma ))P(\Gamma ) \,.  \label{Bc} 
\ee 
Assuming that the dynamics is Markovian, $P(\Gamma )$ is given by: 
\be 
P(\Gamma)=\left[ \prod_{k=0}^{N\sub{s}-1}\prod_{i=1}^{N}q^{k}( \sigma_{i}^{k+1} \mid
 \sigma_{i}^{k} )\right]\prod_{i=1}^{N}p_{o}(\sigma_{i}^{0}) \,,
\label{Bd} 
\ee 
where $q^{k}(\sigma' \mid \sigma)$ is the transition
probability to go from $\sigma$ to $\sigma'$ at time $k$ and $p_{o}$ is
the initial equilibrium distribution.  The details for the computation
of the average work and hopping number are presented in
\ref{NW_C}. Here, we outline the main steps of the computation and give
the final results.

Following \cite{Rit22}, we use the integral representation of the delta
function and write the distribution of work and hopping number as:
$P_{N}(W)=\int d\{ \Gamma \} \rme^{N\cdot a}$ and $P_{N}(M)=\int \rmd\{ \Gamma
\} \rme^{N\cdot b}$, where $a$ and $b$ are functions defined over the space
of trajectories $\{ \Gamma \}$. In the thermodynamic limit
($N\rightarrow \infty$) the problem can be solved by applying the saddle
point technique, i.e. maximizing the functions $a$ and $b$ with respect
to their phase-space variables. From the saddle point equations we
obtain closed expressions for the average quantities, $\langle\dots\rangle $, where the brackets represent an average over an infinite number of
realizations of the ramping protocol.  In the continuous time limit,
corresponding to $\Delta t\rightarrow 0$, $N\sub{s}\rightarrow \infty$ and
$\Delta t \cdot N\sub{s}=t$, the average total work, $\langle W\rangle$,
the average dissipated work, $\langle W_{\rm dis}\rangle$, and the
average hopping number, $\langle M\rangle$, are given
  by (see \ref{NW_C}):
\be
\langle W\rangle &=-\frac{x\sub{m}}{2}\int_{f\sub{min}}^{f\sub{max}}m(f)\rmd f\,, \label{e30} \\
\langle W_{\rm dis}\rangle &=\langle W\rangle -W_{\rm rev}=-\frac{x\sub{m}}{2}\int_{f\sub{min}}^{f\sub{max}}(m(f)-m\sub{eq}(f))\rmd f \,, \label{e3} \\
\langle M\rangle &=\frac{1}{2r}\int_{f\sub{min}}^{f\sub{max}}(m(f)k_{\rm M}(f)+k_{\rm T}(f))(1/r(f))\rmd f\,,
\label{e6}
\ee
where $W_{\rm rev}$ is equal to the reversible work, i.e. the work
measured in the quasi-static limit.  $k_{\rightarrow}(f)$ and
$k_{\leftarrow}(f)$ are the kinetic rates corresponding to the
transitions $\sigma=-1\to 1$ and from $\sigma=1 \to -1$
respectively. $k_{\rm T}(f),k_{\rm M}(f)$ are defined as $k_{\rm
  T}(f)=k_{\rightarrow}(f)+k_{\leftarrow}(f)$, $k_{\rm
  M}(f)=k_{\rightarrow}(f)-k_{\leftarrow}(f)$.  The functions
$m\sub{eq}(f)$ and $m(f)$ are given by:
\be
m\sub{eq}(f)=\frac{k_{\rm M}(f)}{k_{\rm T}(f)} \,, 
\label{e4}
\ee
and
\be
m(f)=m\sub{eq}(f)-\int_{f\sub{min}}^{f}\frac{\rmd m\sub{eq}(f_{1})}{\rmd f_{1}}\exp[-\int_{f_{1}}^{f}\frac{k_{\rm T}(f_{2})}{r(f_{2})}df_{2}]\rmd f_{1}\,,
\label{e5}
\ee
where $r$ is the loading rate defined as $r=\frac{\rmd f(t)}{\rmd t}$. 

\subsection{Constant loading rate}
\label{constantrate}
From now on we consider the case in which the 
force $f$ is varied from $f\sub{min}$ to $f\sub{max}$ at a constant loading rate $r$. 
The initial and final values of the force, $f\sub{min}$ and $f\sub{max}$, are such that the 
equilibrium probabilities to be in the F state ($\sigma=-1$) are equal 1 and 0 respectively, i.e. 
$m\sub{eq}(f\sub{min})=-1$ and $m\sub{eq}(f\sub{max})=1$. Note that such condition 
is always experimentally verified along the stretching process where initially the
molecule is always folded and finally unfolded at the end of the
process.

Interestingly, when the dynamics is governed by the rates in \eref{e7}, 
$\langle W_{\rm dis}\rangle/k\sub{B}T$ and $\langle M\rangle$ become functions 
of only two adimensional parameters, the fragility $\mu$ and an
adimensional rate $\tilde{r}$:
\be
\fl \frac{\langle W_{\rm dis}\rangle}{k\sub{B}T} &=&\int_{-\infty}^{\infty}\rmd x \int_{-\infty}^{x}\rmd y\,\frac{1}{{\rm{cosh}}^{2}y}\exp \left(-\frac{1}
{\tilde{r}}\int_{y}^{x}\rmd z\,\rme^{\mu z}\cosh z\right)\,,
\label{en1} \\
\fl \langle M\rangle &=&\frac{1}{2 \tilde{r}}\left[ \int_{-\infty}^{\infty}\rmd x (1-\tanh^{2}x)\rme^{\mu x}\cosh x\right.\nonumber\\
\fl &&\left.+\int_{-\infty}^{\infty}\rmd x\, \rme^{\mu x}\sinh x \int_{-\infty}^{x}\rmd y\frac{1}{{\rm{cosh}}^{2}y}\exp \left(-\frac{1}
{\tilde{r}}\int_{y}^{x}\rmd z\,\rme^{\mu z}\cosh z\right)\right]\,.
\label{en2}
\ee
The fragility $\mu$ has been defined in \eref{mu} whereas the adimensional rate $\tilde{r}$ is given by
\be
\tilde{r}=\frac{x\sub{m}r}{k\sub{c}4k\sub{B}T}\,,
\label{adrate}
\ee
$k\sub{c}$ being the folding-unfolding rate at the coexistence force value $f^{\rm c}$ at
which the F and UF states are equally populated,
$k\sub{c}=k_{\rightarrow}(f\sub{c})=k_{\leftarrow}(f\sub{c})$ . \Eref{en1} is
the equivalent of equation (49) in \cite{Rit22} obtained for the
Glauber kinetics of a magnetic dipole in a magnetic field.

Since our goal is to identify relations between measurable quantities
and the parameters characterizing the molecule, we have investigated
the loading rate dependence of $\langle M\rangle$ and $\langle W_{\rm
dis}\rangle$. In \fref{fig4} we show the average
dissipated work and the average hopping number as a function of the
adimensional rate $\tilde{r}$ for different fragilities $\mu$ obtained by
numerical integrating \eref{en1} and \eref{en2}.

\subsection{The stretching versus the releasing processes} 
\label{F-R}
Let us consider a symmetric protocol, i.e. in the stretching process the
force $f$ increases from $f\sub{min}$ to $f\sub{max}$ at a loading rate $r$, while in
the releasing process $f$ decreases from $f\sub{max}$ to $f\sub{min}$ at the same
rate. According to \eref{en1} and \eref{en2} $\langle W_{\rm dis}\rangle$ and $\langle
M\rangle$ are sole functions of $\tilde{r},\mu$. The transformation from
the stretching to the releasing force protocol (stretching/release
transformation) corresponds to the exchange $x^{\rm F}\rightleftharpoons
x^{\rm UF}$ implying ${\mu}\rightarrow -{\mu}$ in
\eref{en1} and \eref{en2}. Moreover, since $r$ and $x\sub{m}$ change sign under the
stretch/release transformation, the parameter $\tilde{r}$ is invariant
under such transformation. For a given
value of $\tilde{r}$ we can write
\be 
\langle W_{\rm dis}\rangle\sub{S}(\mu)=\langle W_{\rm
dis}\rangle\sub{R}(-\mu)\,, \qquad \langle M\rangle\sub{S}(\mu)=\langle
M\rangle\sub{R}(-\mu)\,.
\label{sym1}
\ee
Moreover it is easy to prove that the average hopping number
  $\langle M\rangle$ \eref{en2} is invariant under the transformation
  $\mu\to -\mu$, whereas the average dissipated work $\langle W_{\rm
    dis}\rangle$ \eref{en1} is not.  $\langle M\rangle(\mu)$ verifies:
\be
\langle M\rangle(\mu)\equiv \langle M\rangle\sub{S}(\mu)=\langle M\rangle\sub{S}(-\mu)=\langle
M\rangle\sub{R}(-\mu)=\langle M\rangle\sub{R}(\mu)\,.
\label{sym2}
\ee
This is a symmetry relation that can be checked in simulations and experiments. Results for  $\langle W_{\rm dis}\rangle$ and $\langle M\rangle$ for different values of $\mu$ are shown in the top panels of \fref{fig4}.

In \cite{ManRit05} (see the section {\it Fraction of trajectories that have at
least one refolding} and the Appendix C) it was proven that the average
fraction $\phi$ of trajectories (a quantity different from $M$) with more than one unfolding
(refolding) (i.e. with $M\ge3$) is equal along the stretching and release processes for
a symmetric perturbation protocol. Although we have been unable to
compute such quantity $\phi$ using path thermodynamics it is interesting to
see that the same symmetry relation \eref{sym2} is also satisfied by
the average hopping
number $\langle M\rangle$. 

In the numerical simulations done in \cite{ManRit05} (see figure 11 in
that reference) it was also shown that the fraction $\phi$, when plotted
as a function of the average dissipated work along the stretching
process, $\langle W_{\rm dis} \rangle\sub{S}$, collapse into a single curve
for different two-state systems (i.e. that are characterized
by different values of the parameters $\Delta G$, $x^{\rm F}$, $x^{\rm
  UF}$ and $k\sub{m}$).  Although we cannot verify or falsify such relation
for $\phi$, we can see that \eref{sym2} precludes the validity of a similar relation for
$\langle M\rangle$. As shown in the bottom left panel in
\fref{fig4}, the relation between $\langle W_{\rm dis} \rangle\sub{S}$ and
$\langle M\rangle$ is not unique for different values of $\mu$
(otherwise $\langle W_{\rm dis} \rangle\sub{S}(\mu)$ would be identical to
$\langle W_{\rm dis} \rangle\sub{R}(\mu)$, which we know it is not
  generally true).

By applying \eref{en1} to the stretching and releasing processes it is possible
to isolate the variable $\tilde{r}$ and express $\langle W_{\rm
dis}\rangle\sub{S}$ as a function of $\langle W_{\rm dis}\rangle\sub{R}$ and
the fragility $\mu$. By plotting $\langle W_{\rm dis}\rangle\sub{S}$ as a
function of $\langle W_{\rm dis}\rangle\sub{R}$ for different values of
$\mu$ we can get {\it isofragility} curves that characterize the
dissipation of the molecule along the stretching and releasing
processes. The results are shown in the bottom right panel of 
\fref{fig4}. The relation between the dissipated work
measured in the stretching and releasing processes can
be then used to obtain information about $\mu$, and hence about the
TS of the folding/unfolding reaction.

\subsection{Low loading rate regime}
\label{lowr}
\begin{figure}
\begin{center}
\vspace{0.9cm}
\includegraphics[width=14.5cm,angle=0]{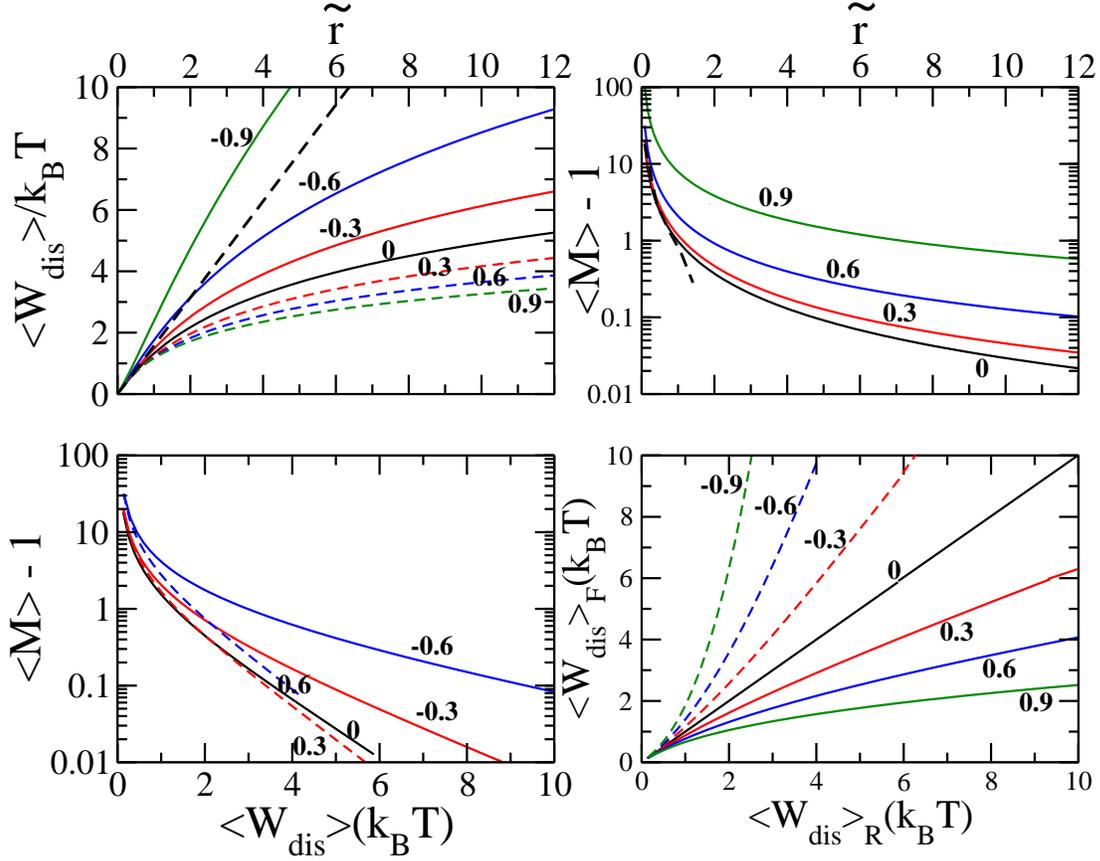}

\caption{Various plots of $\langle W_{\rm
dis}\rangle$ (in $k\sub{B}T$ units) and $\langle M\rangle$ as a function of the adimensional
loading rate $\tilde{r}$ obtained by numerical integrating 
\eref{en1} and \eref{en2} for various values of $\mu$ (indicated in the panels). (Top
left) $\langle W_{\rm dis}\rangle\sub{S}$ along the stretching 
process as a function of $\tilde{r}$. (Top right) $\langle
M\rangle=\langle M_F\rangle=\langle M_R\rangle $ as a function of
$\tilde{r}$ for both the stretching and releasing processes. The dashed line
corresponds to the expansion \eref{expan2b} for $\mu=0$. (Bottom left)
$\langle M\rangle$ as a function of $\langle W_{\rm dis}\rangle\sub{S}$ along
the stretching process. (Bottom right) $\langle W_{\rm dis}\rangle\sub{S}$ as a
function of $\langle W_{\rm dis}\rangle\sub{R}$. Note that $\langle W_{\rm
dis}\rangle\sub{S}<\langle W_{\rm dis}\rangle\sub{R}$ for $\mu>0$,  $\langle W_{\rm
dis}\rangle\sub{S}>\langle W_{\rm dis}\rangle\sub{R}$ for $\mu<0$ whereas for
$\mu=0$ (TS located in the middle of the two states) $\langle W_{\rm
dis}\rangle\sub{S}=\langle W_{\rm dis}\rangle\sub{R}$.}
\label{fig4}

\end{center}
\end{figure}

Equations \eref{en1} and \eref{en2} are not analytically integrable, however we
can solve them numerically as well as study their asymptotic
behaviour in the low loading rate regime. The low loading rate regime is
investigated by expanding the expressions for $\langle W_{\rm
dis}\rangle/k\sub{B}T$ and $\langle M\rangle$, \eref{en1} and \eref{en2}, around $\tilde{r}=0$. The
analytical computation of the different terms of the expansion is
presented in \ref{NW_D}. We report the final result:
\be
\frac{\langle W_{\rm dis}\rangle}{k\sub{B}T}&=&\frac{\pi}{2}(1-\mu^{2})\sec (\pi\mu/2)\tilde{r}-\frac{2}{3}\mu^{2}(1-\mu^{2})\pi\csc(\pi\mu)\tilde{r}^{2}+\nonumber \\
&+& \frac{3}{40}(-5+51\mu^{2}-55\mu^{4}+
+9\mu^{6})\pi\sec(3\pi\mu/2)\tilde{r}^{3}+\Or(\tilde{r}^{4})\,,
\label{e9} \\
\langle M\rangle&=&(\pi/2)\sec (\pi\mu/2)\frac{1}{\tilde{r}}+\frac{1}{48}(9-10\mu^{2}+\mu^{4})\pi\sec(\pi\mu/2)\tilde{r}\nonumber\\
&-&\frac{1}{3}(\mu-\mu^{3})\pi\csc(\pi\mu)\tilde{r}^{2}+\Or (\tilde{r}^{3})\,.
\label{e10}
\end{eqnarray}
The first term in the expansion of $\langle W_{\rm dis}\rangle$ agrees with the
linear response result previously reported in \cite{Rit1} for the case
$\mu=\to -1$ (i.e. $x^{\rm F}\to 0$, corresponding to an unfolding rate
independent of force). For that case we get
\be
\frac{\langle W_{\rm dis}\rangle}{k\sub{B}T}=2\tilde{r}+\Or (\tilde{r}^2)\label{expan1a}\\
\langle M\rangle\to \infty\,,\label{expan1b}
\ee
whereas for the case $\mu=0$ (corresponding to a TS located in the middle between the F
and UF states) we get
\be
\frac{\langle W_{\rm dis}\rangle}{k\sub{B}T}=\frac{\pi}{2}\tilde{r}+\Or (\tilde{r}^2)\label{expan2a}\\
\langle M\rangle=\frac{\pi}{2\tilde{r}}+\Or (\tilde{r})\,. \label{expan2b}
\ee
In \fref{fig4} (top panels) we plot (black dashed lines) the expansion
\eref{expan2a} (\fref{fig4}, top left panel) and the expansion \eref{expan2b}
(\fref{fig4}, top right panel) for the case $\mu=0$.

\section{Dissipation in the mixed ensemble} 
\label{renorm}
In the previous analysis we considered that the force is externally
controlled and does not fluctuate (force ensemble). However in the
optical tweezers experiments this is not the case because it is the
position of the optical trap relative to the pipette (rather than the
force) that is controlled. This corresponds to what has been called
the mixed ensemble (see our companion paper \cite{MosManForHugRit08}).
The kinetic rates measured in the mixed ensemble are different that
those measured in the force ensemble
\cite{GerBunHwa01,GerBunHwa03,ManRit05,WenManLiSmiBusRitTin07,ManWenLiSmiBusTinRit07}.
This is a consequence of the extrem sensitivity of the kinetic rates
on height of the kinetic barrier (through the Arrhenius exponential
dependence). Just a $\frac{1}{2}$ kcal/mol variation at room
temperature in the value of the kinetic barrier can modify the rates
by a factor of 2 (100\%). Because the value of the kinetic rate
$k\sub{c}$ enters into the the definition of the adimensional rate
$\tilde{r}$ \eref{adrate}, a 100\% variation in its value will
invalidate an adequate comparison between theory and experiment.  The
influence of the ensemble on the kinetics is yet another example of
nonequilibrium thermodynamics applied to small systems
\cite{BusLipRit05}, i.e. about how nonequilibrium effects strongly
depend on which experimental variables are controlled when the size of
the system is small enough.

In the current theory the dissipated work and the average hopping
number were obtained in the force ensemble. A path thermodynamics
calculation for the mixed ensemble seems too tedious for the same
analysis to be repeated again. In what follows we show that it is
possible to carry over the theory in the mixed ensemble into the
force-ensemble theory developed in \sref{Ana} just by appropriate
rescaling of the kinetic rates \eref{e7} by a rescaling factor
$\Omega$. This rescaling only affects the value of the critical rate
$k\sub{c}$ that enters into the definition of $\tilde{r}$
\eref{adrate} whereas the rest of parameters (such as $x\sub{m}$ and
$\mu$) remain unchanged.

\subsection{A simple model for the mixed ensemble}
\label{simplemodel}
\begin{figure}
\begin{center}
\includegraphics[width=16cm,angle=0]{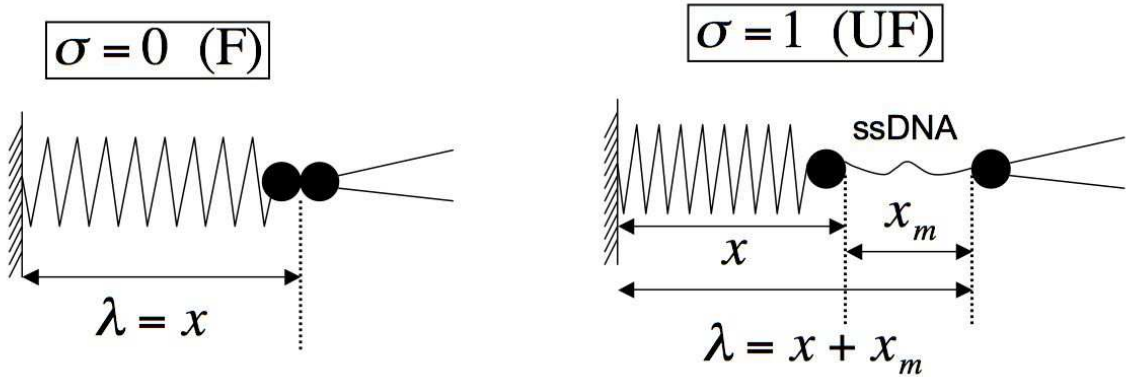}

\caption{ Schematic picture of a simple model for the mixed
    ensemble. (Left) Folded state $\s=0$. (Right) Unfolded state
    $\s=1$. See main text for the explanation of the different
    parameters.}
\label{fig5}

\end{center}
\end{figure}

Let us consider the system formed by a single hairpin in series with a
Hookean spring (\fref{fig5}). The spring has stiffness equal to $k\sub{x}$ whereas the molecule can be in two
states, $\s=0$ (folded) and $\s=1$ (unfolded). This Hookean spring
stands for the combined effect of the bead in the optical trap and the
handles used to manipulate the hairpin. The total extension of the system is
\be
\lambda=x+x\sub{m}\sigma \,,
\label{ap0}
\ee
 where $x$ is the extension of the spring and $x\sub{m}$
is the molecular extension of the unfolded molecule. We will assume that
$x\sub{m}$ is not force dependent and we will take a zero extension
for the folded molecule.

In the mixed-ensemble protocol the total distance $\lambda$ (equal to the distance between the optical trap and the micropipette) is controlled. The total energy of the system can be expressed as
\be
E(\lambda,\sigma)=\frac{1}{2}k\sub{x} (\lambda-x\sub{m} \sigma)^2+\sigma \Delta G \,,
\label{ap1}
\ee
where $\Delta G$ is the free energy difference between the unfolded and
the folded states of the hairpin. The force-distance curve (FDC) is given by
\be
f_{\s}(\lambda)=\frac{\partial
E(\lambda,\s)}{\partial \lambda}=k\sub{x}(\lambda-x\sub{m} \sigma)=k\sub{x} x\,.
\label{ap2}
\ee
The FDC of the system has two branches depending on whether the molecule is folded or unfolded. If the molecule is folded ($\sigma=0$) we have
\be
f\sub{F}(\lambda)=\frac{\partial E(\lambda,0)}{\partial \lambda}=k\sub{x}\lambda\,,
\label{ap2a}
\ee
whereas in the other case,
\be
f\sub{UF}(\lambda)=\frac{\partial E(\lambda,1)}{\partial \lambda}=k\sub{x}(\lambda-x\sub{m}) \,,
\label{ap2b}
\ee
i.e both force-distance curves have the same slope ($k\sub{x}$), but there is
a drop in the force equal to $k\sub{x} x\sub{m}$ when the molecule
unfolds. Note that this is exactly what is observed in the
force-extension curves shown in \fref{fig2}.

What is the appropriate theory for the dissipated work in the mixed
ensemble where the total distance ($\lambda$) rather than the force ($f$)
is controlled? In principle we should expect work fluctuations to differ
in both ensembles. 
The mechanical work along a trajectory $\Gamma$ in the force and mixed ensemble, $W_{f}$ and $W_{\lambda}$ respectively, 
is given by
\be
W_f(\Gamma)&=-x\sub{m}\int_{f\sub{min}}^{f\sub{max}} \s \rmd f\,, \label{ap4} \\
W_\lambda(\Gamma)&=\int_{\lambda\sub{min}}^{\lambda\sub{max}} f \rmd\lambda\,.
\label{ap5a}
\ee
$f\sub{min},f\sub{max}$ ($\lambda\sub{min},\lambda\sub{max}$) are the initial and final values of the force (total distance), 
 the trajectory $\Gamma$ is defined by the time evolution of $\lbrace \s(t)\rbrace$, and 
\be
x\sub{m}=x^{\rm F}+x^{\rm UF}\,.
\label{ap3c}
\ee

Integrating by parts $W_{\lambda}$ and using the relations \eref{ap2}
and \eref{ap0}, we get
\be
W_\lambda(\Gamma)&=[f(\lambda\sub{max})\lambda\sub{max}-f(\lambda\sub{min})\lambda\sub{min}]-\nonumber\\&\frac{1}{2k\sub{x}}[f^2(\lambda\sub{max})^2-f^2(\lambda\sub{min}))-x\sub{m} \int_{f(\lambda\sub{min})}^{f(\lambda\sub{max})} \s
\rmd f=\nonumber\\
&=W_f(\Gamma)+(f\sub{max}\lambda\sub{max}-f\sub{min}\lambda\sub{min})-\frac{1}{2k\sub{x}}(f\sub{max}^2-f\sub{min}^2)\,,
\label{ap5b}
\ee
where in the last line  we have taken the instantaneous
  forces in the mixed ensemble equal to their values in the force ensemble, $f\sub{min}=f(\lambda\sub{min}),f\sub{max}=f(\lambda\sub{max})$. Although work distributions are
  expected to differ in both ensembles, differences in the average dissipated work are expected
  to be negligible. Therefore, $\langle W_\lambda^{\rm
diss}(\Gamma)\rangle=\langle W_f^{\rm
diss}(\Gamma)\rangle$ with $W_\lambda^{\rm
diss}(\Gamma)=W_\lambda(\Gamma)-W_\lambda^{\rm rev},W_f^{\rm
diss}(\Gamma)=W_f(\Gamma)-W_f^{\rm rev}$. 
If the average dissipated work is the same
in both ensembles then where
is the expected difference between the measured dissipated work in the
two experimental conditions (controlled $f$ versus controlled
$\lambda$)?. The answer is in the kinetics. Because the kinetic rates in both
ensembles are different, the ensemble of paths $\Gamma$ generated in
both ensembles have different work and dissipated work distributions.

\subsection{Rescaling factor for the kinetic rates}
\label{renorm_factor}
Let us define $K_{\rightarrow}(\lambda),K_{\leftarrow}(\lambda)$ as the
rates in the mixed ensemble (we use major case letters to distinguish
them from the force-ensemble rates). The
mixed-ensemble rates must satisfy the detailed balance condition,
\be
\frac{K_{\rightarrow}(\lambda)}{K_{\leftarrow}(\lambda)}=\exp\left(-\beta(E(\lambda,1)-E(\lambda,0))\right)\,.
\label{ap6}
\ee
Inserting the expression for the energy \eref{ap1} we get
\be
\frac{K_{\rightarrow}(\lambda)}{K_{\leftarrow}(\lambda)}=
\exp\left(-\beta(\frac{1}{2}k\sub{x}((\lambda-x\sub{m})^2-\lambda^2)+\Delta G)\right)\,.
\label{ap6a}
\ee
We can now express the difference of the squared terms in \eref{ap6a} as a product of a
sum times a difference of two terms. However, according to
the FDCs for the two branches \eref{ap2a} and \eref{ap2b}, the factor 1/2 times the sum is
just the $\lambda$-dependent average force $\overline{f}$ between the two branches, 
\be
\overline{f}(\lambda)=\frac{1}{2}(f\sub{F}(\lambda)+ f\sub{UF}(\lambda))=\frac{1}{2}k\sub{x}(2\lambda-x\sub{m})\,.
\label{ap6b}
\ee
Therefore,
\be
\frac{K_{\rightarrow}(\overline{f})}{K_{\leftarrow}(\overline{f})}=
\exp[-\beta(-\overline{f} x\sub{m}+\Delta G)]\,.
\label{ap6c}
\ee
where we parametrize the rates 
in the mixed ensemble $K_{\rightarrow}(\lambda),K_{\leftarrow}(\lambda)$
in terms of $\overline{f}$ rather than $\lambda$. Because these rates
must satisfy detailed balance \eref{ap6c} we can write 
\be
K_{\rightarrow}(\overline{f})=k_{\rightarrow}(\overline{f}+a) \,,
\label{ap61a}\\
K_{\leftarrow}(\overline{f})=k_{\leftarrow}(\overline{f}+b)\,, \label{ap61b}
\ee
where $k_{\rightarrow},k_{\leftarrow}$ are the kinetic rates
  \eref{e7} and $a,b$ are two arbitrary functions of $\overline{f}$
(i.e. of $\lambda$). Using the definitions \eref{e7} and inserting them
in \eref{ap6c}, we obtain the identity
\be
a x^{\rm F}+b x^{\rm UF}=0\,.
\label{ap62}
\ee
There is an infinite number of possible solutions that satisfy this
equation. The most general solution is
\be
a=C x^{\rm UF}\,, \qquad b=-C x^{\rm F} \,, 
\label{ap63}
\ee
with $C$ an arbitrary function of $\lambda$. We now express $a,b$ in terms of the fragility $\mu$
defined in \eref{mu}. It is straightforward to verify that
\be
x^{\rm F}=\frac{\mu+1}{2}x\sub{m}\,, \qquad x^{\rm UF}=\frac{1-\mu}{2}x\sub{m}
\label{ap65}
\ee
where we used \eref{ap3c}. We now introduce the difference $\Delta f$
between the forces along the F and UF branches at a
given value of $\lambda$,
\be
\Delta f=f\sub{F}(\lambda)-f\sub{UF}(\lambda)=k\sub{x} x\sub{m}\,.
\label{ap66}
\ee
Note that in this simple model $\Delta f$ is independent of $\lambda$
(incidentally, let us note that the difference of
force between the two branches shown in \fref{fig2} is approximately constant). Inserting
\eref{ap65} and \eref{ap66} into \eref{ap63}, we get
\be
a=\alpha (1-\mu)\Delta f\,, \qquad b=-\alpha (1+\mu)\Delta f \,,
\label{ap67}
\ee
where $\alpha$ is a dimensionless constant. Substituting \eref{ap67} into
\eref{ap61a} and \eref{ap61b}, we finally obtain 
\be
K_{\rightarrow}(\overline{f})=k_{\rightarrow}\left(\overline{f}+\alpha(1-\mu)\Delta
f\right)\,,\label{ap7a}\\
K_{\leftarrow}(\overline{f})=k_{\leftarrow}\left(\overline{f}-\alpha(1+\mu)\Delta
f\right)\,,\label{ap7b} 
\ee
where $\alpha$ remains undetermined.

We cannot proceed further unless we introduce a microscopic model for
the folding/unfolding of the hairpin in the proper experimental
setup. This was done in \cite{ManRit05} where Kramers theory was
applied to investigate thermodynamic and kinetic aspects of RNA
hairpins in the mixed ensemble relevant to pulling experiments using
optical tweezers. The expression for the kinetic rates were obtained
in formulae (B5), (B6) in the Appendix B of that reference. These
satisfy \eref{ap7a} and \eref{ap7b} with $\alpha=1/4$. If we insert $\alpha=1/4$
in \eref{ap7a} and \eref{ap7b} and use \eref{e7}, we get
\be
K_{\rightarrow}(\overline{f})=k_{\rightarrow}(\overline{f})
\exp\left(\beta\frac{(1-\mu)}{4}x^{\rm F}\Delta f\right)\,,\label{ap8a}\\
K_{\leftarrow}(\overline{f})=k_{\leftarrow}(\overline{f})
\exp\left(\beta\frac{(1+\mu)}{4}x^{\rm UF}\Delta f\right)\,.\label{ap8b}
\ee
Now we use again \eref{ap65} to obtain
\be
k_{\rightarrow}(\overline{f})=\Omega K_{\rightarrow}(\overline{f})\,,\label{ap9a}\\
k_{\leftarrow}(\overline{f})=\Omega K_{\leftarrow}(\overline{f})\,,\label{ap9b}
\ee
where $\Omega$ is a rescaling factor for the kinetic rates defined as
\be
\Omega=\exp\left[-\left(\beta\frac{1-\mu^2}{8}x\sub{m}\Delta f\right)\right] \,,
\label{ap10}
\ee
showing that both rates  $K_{\rightarrow}, K_{\leftarrow}$ must be rescaled
with the same factor $\Omega$.

\subsection{The critical rate in the mixed ensemble, $\overline{k}_c$.}
\label{criticalmixed}
We can now extend the theory developed for the force ensemble to the
mixed ensemble. Instead of the simplified rates \eref{e7} we must use the rescaled rates
\eref{ap9a} and \eref{ap9b}. Consequently, all the results obtained in
\sref{Ana} hold but by using the rates \eref{ap9a} and \eref{ap9b} instead of \eref{e7}. 

The value of the rates $K_{\rightarrow}(\overline{f}),
K_{\leftarrow}(\overline{f})$ in the mixed ensemble can be measured in
hopping experiments in the passive mode
\cite{WenManLiSmiBusRitTin07,ManWenLiSmiBusTinRit07}. In these experiments 
 the total extension $\lambda$ was held  fixed and the evolution of the force is recorded.
 A typical passive mode force trajectory shows a square-like signal in 
which the force jumps between two values (see below in Figure~\ref{fig6})
, the high value $f\sub{F}$ corresponding to the F state and the low one $f\sub{UF}$ 
corresponding to the UF state. Two types of
  representation for the passive rates were adopted in these
  references. In one representation rates are plotted as functions of
  $\overline{f}$ (or $\lambda$). These were called apparent rates,
  $K_{\rightarrow}^{\rm app}(\overline{f}), K_{\leftarrow}^{\rm app}(\overline{f})$,
  to distinguish them from {\it plain} passive rates where the relevant
  variable is the force corresponding to the folded or unfolded
  branches, $f\sub{F},f\sub{UF}$, $K_{\rightarrow}^{\rm plain}(f\sub{F}), K_{\leftarrow}^{\rm plain}(f\sub{UF})$. The rates
$K_{\rightarrow}, K_{\leftarrow}$ introduced in the previous section
correspond to the apparent rates measured in passive hopping. 

The value of the coexistence rate $\overline{k}\sub{c}$ that must be used in the
definition of $\tilde{r}$ in \eref{adrate} is given by
\be
\overline{k}\sub{c}=k_{\rightarrow}(\overline{f}\sub{c})=k_{\leftarrow}(\overline{f}\sub{c})
\label{ap11} 
\ee 
and using the results \eref{ap9a} and \eref{ap9b} we have
\be
\overline{k}\sub{c}=\Omega K\sub{c}^{\rm app} \,,
\label{ap12}
\ee
where $K\sub{c}^{\rm app}$ is the critical apparent rate in the mixed ensemble defined as
\be
K\sub{c}^{\rm app}=K_{\rightarrow}^{\rm app}(\overline{f}\sub{c})=K_{\leftarrow}^{\rm app}(\overline{f}\sub{c})\,,
\label{ap13}
\ee
where from \eref{ap7a}, \eref{ap7b} and \eref{e7} we obtain $\overline{f}\sub{c}=f\sub{c}=\Delta G_1/x\sub{m}$. 

By plotting the passive apparent rates as a function of
$\overline{f}$ we can determine $K\sub{c}^{\rm app}$ as the
value of the rate at which  $K_{\rightarrow}^{\rm
    app}(\overline{f})$ and
$K_{\leftarrow}^{\rm app}(\overline{f})$ intersect
each other. The plain and the apparent rates satisfy
\be
K_{\rightarrow}^{\rm app}(\overline{f})=K_{\rightarrow}^{\rm
plain}(\overline{f}+\frac{\Delta f}{2})\,,\label{ap14a}\\
K_{\leftarrow}^{\rm app}(\overline{f})=K_{\leftarrow}^{\rm
plain}(\overline{f}-\frac{\Delta f}{2})\,,\label{ap14b}
\ee
and from \eref{ap9a} and \eref{ap9b} we obtain
\be
k_{\rightarrow}(\overline{f})=\Omega K_{\rightarrow}^{\rm
plain}(\overline{f}+\frac{\Delta f}{2})\,,\label{ap15a}\\
k_{\leftarrow}(\overline{f})=\Omega K_{\leftarrow}^{\rm
plain}(\overline{f}-\frac{\Delta f}{2})\,.\label{ap15b}\\ 
\ee
Finally, it is easy to verify that $K\sub{c}^{\rm plain}=K\sub{c}^{\rm app} \Omega^2$ and from \eref{ap12}
we get
\be
\overline{k}\sub{c}=\frac{K\sub{c}^{\rm plain}}{\Omega}\,.
\label{ap16}
\ee
Because $\Omega<1$ we have the following chain of inequalities:
\be
K\sub{c}^{\rm plain} < \overline{k}\sub{c} < K\sub{c}^{\rm app}\,.
\label{ap17}
\ee
Note that all these three rates are defined in the mixed ensemble and
cannot be directly compared to experimental rates measured in
hopping experiments carried out in the constant force mode (i.e. in the force
ensemble). The value of critical rate measured in hopping experiments in
the constant force mode (what we called $k\sub{c}$ in \sref{2-s} and is used in
\eref{en1} and \eref{en2}) should be compared to pulling measurements of $\langle
W_{\rm diss}\rangle$ and $\langle M\rangle$ done in the force ensemble, where the force
acting on the bead (rather than the position of the trap) is controlled
and ramped at a constant rate.

\section{Experimental results}
\label{expresults}
In this section we compare the experimental results for $\langle
W_{\rm dis}\rangle$ and $\langle M \rangle$ that we obtained in
pulling experiments for the DNA hairpin of \fref{fig1}(b) with the
theoretical predictions developed in the preceding sections. Important
parameters to compare theory and experiments are: the distances
$x^{\rm F}$ and $x^{\rm UF}$, necessary to determine the fragility
$\mu$ \eref{mu}, and the coexistence rate, $\overline{k}\sub{c}$, necessary to
determine the adimensional rate \eref{adrate}.  A detailed
characterization of the kinetics of a hairpin can be done with pulling
or hopping experiments. In our companion paper
\cite{MosManForHugRit08} we investigated in detail the
folding/unfolding kinetics of the hairpin using pulling
experiments. However, accurate estimates for $\overline{k}\sub{c}$ are
easier to obtain in hopping experiments. As we explained in
\sref{criticalmixed} the hopping experiments should be carried out
in the passive mode
\cite{WenManLiSmiBusRitTin07,ManWenLiSmiBusTinRit07}. All experiments
were done at a temperature $23^{\circ}-24^{\circ}$ C in a 1M NaCl aqueous
buffer with neutral pH (7.5) stabilized by Tris HCl and 1M
EDTA.

\Table{\label{table1} Collected statistics from pulling data. The parameteres are: $v$ (pulling
   speed in nm/s); $r$ (loading rate in pN/s); $N\sub{M}$ (total number of molecules); $N\sub{S}$
   (total number of stretchings); $N\sub{R}$   (total number of releasings);
   $\langle W_{\rm dis}\rangle\sub{S}$ (average dissipated work along
   the stretching process in $k\sub{B}T$ units); $\langle W_{\rm dis}\rangle\sub{R}$ (average dissipated work along
   the releasing process in $k\sub{B}T$ units); $\langle M\rangle\sub{S}$ (average
   hopping number  along the stretching process); $\langle M\rangle\sub{R}$ (average
   hopping number  along the releasing process). The experimental errors
   are shown in parenthesis and give an estimate of the
   variability among different molecules. (*) For the slowest pulls (25
   nm/s) we have data for just one molecule so we do not indicate the
   experimental error.}
\br
$\0\0v$&$\0r$&$N\sub{M}$&$N\sub{S}$&$N\sub{R}$&$\langle W_{\rm dis}\rangle\sub{S}$&$\langle W_{\rm
  dis}\rangle\sub{R}$&$\langle M\rangle\sub{S}$&$\langle M\rangle\sub{R}$ \\
 \mr
\018.5& \01.0  & 1 & 193 & 192 & 1.14(*)   & 1.00(*)   &  3.21(*)&3.13(*)   \\
\036.5& \01.95& 2 & 175 & 174 & 2.1(2)& 1.80(1)&  1.66(13)  &1.70(9) \\
\086.2& \04.88& 3 & 567 & 565 & 3.03(11)& 3.41(12)& 1.17(3)&1.11(3) \\
156    & \08.1   & 2 & 721 & 723 & 3.95(17)& 4.29(2) & 1.047(16)&1.080(6)  \\
274.3& 14.9    & 2 & 827 & 806 &  4.72(20)& 5.5(5)    & 1.01(1)&1.04(2)   \\
\br
\endTable

\subsection{Passive hopping experiments} 
\label{hopping}
\begin{figure}
\begin{center}
\includegraphics[width=9cm,angle=0]{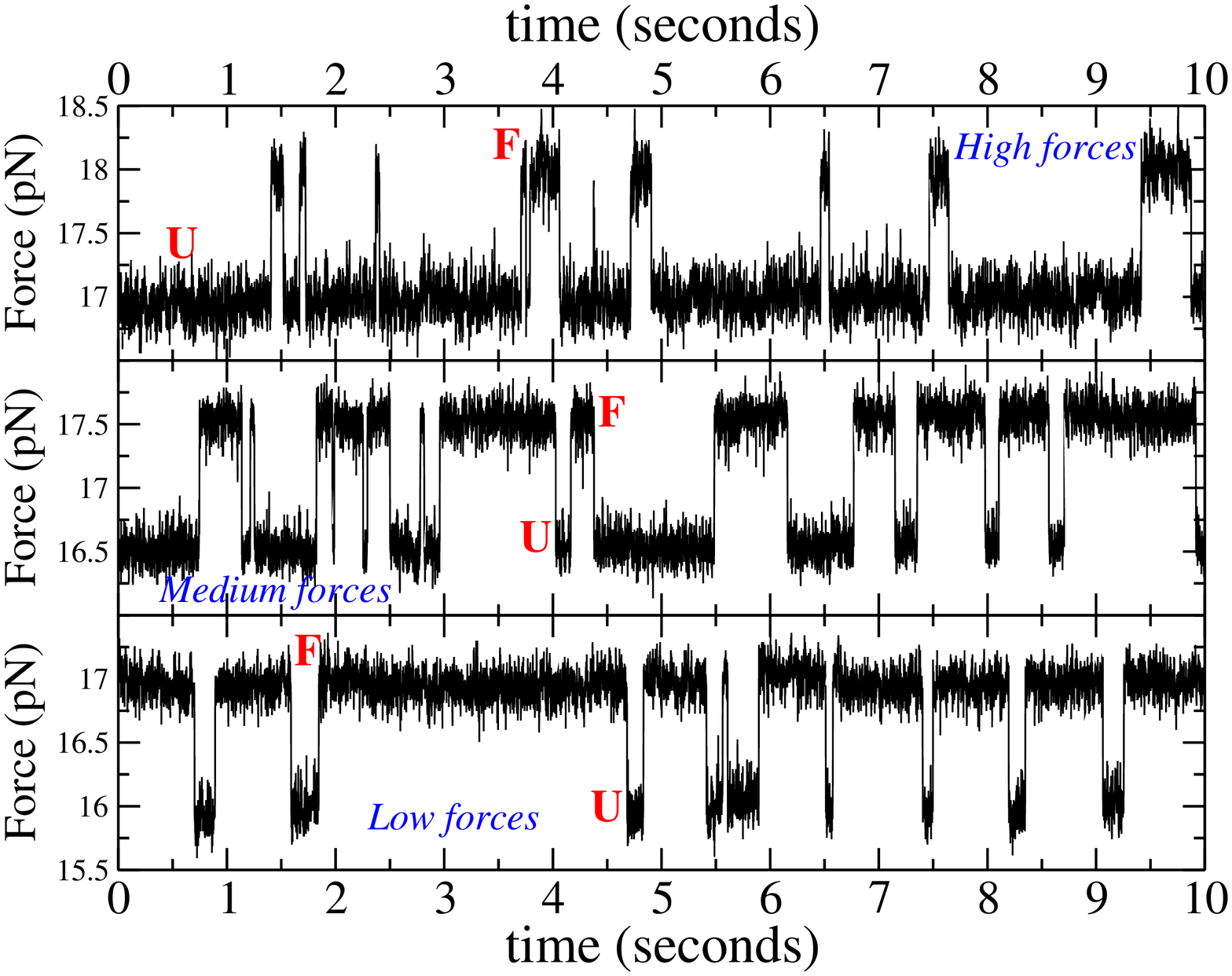}
\includegraphics[width=9cm,angle=0]{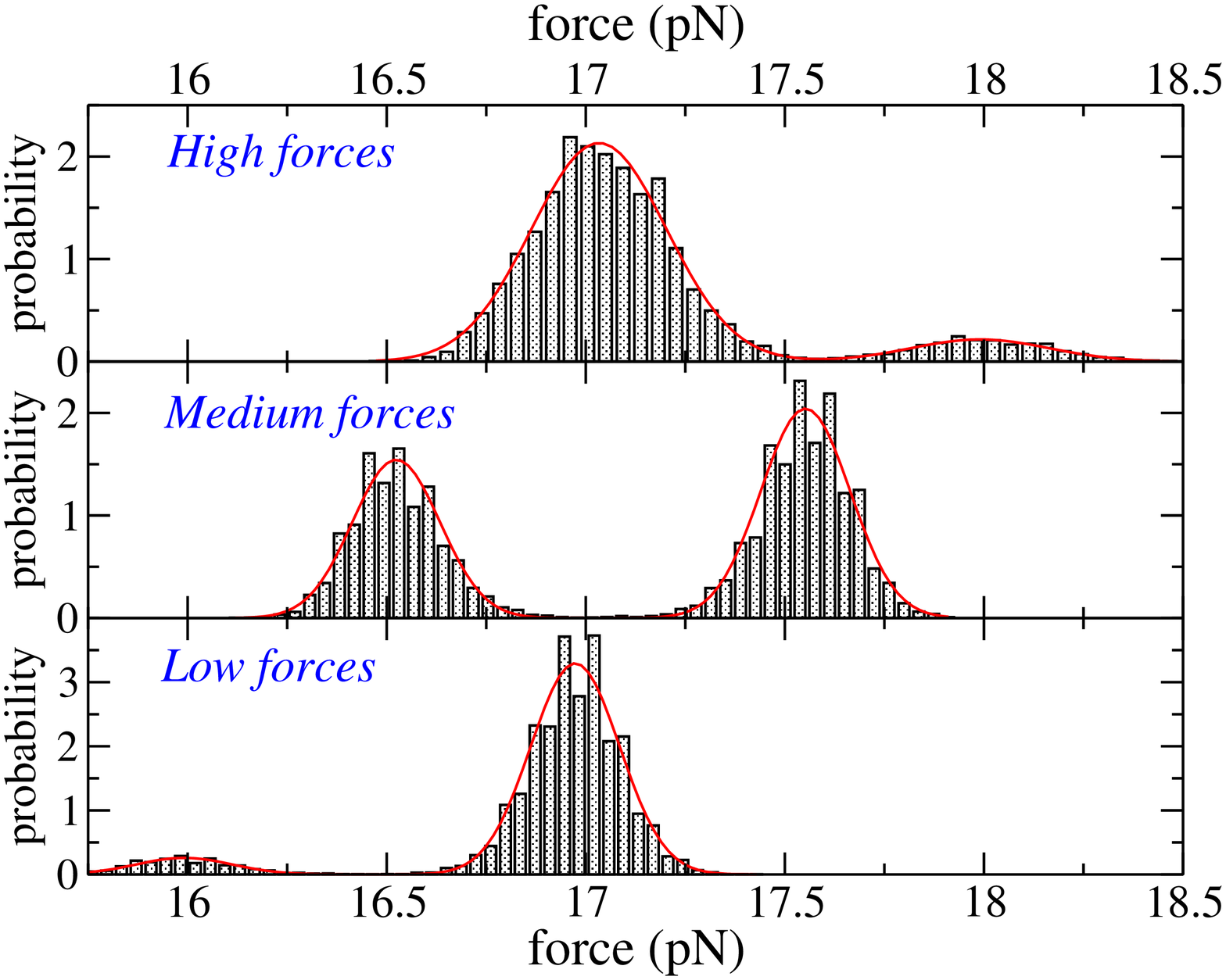}
\caption{Hopping experiments in the passive mode.
(Upper panel). A typical hopping trace measured at three trap positions: upper (high forces),
middle (intermediate forces) and bottom (lower forces). (Lower panel)
Force distributions for the dichotomic signal shown in the upper panel.}
\label{fig6}
\end{center}
\end{figure}

\begin{figure}
\begin{center}
\vspace{0.9cm}
\includegraphics[width=10cm,angle=0]{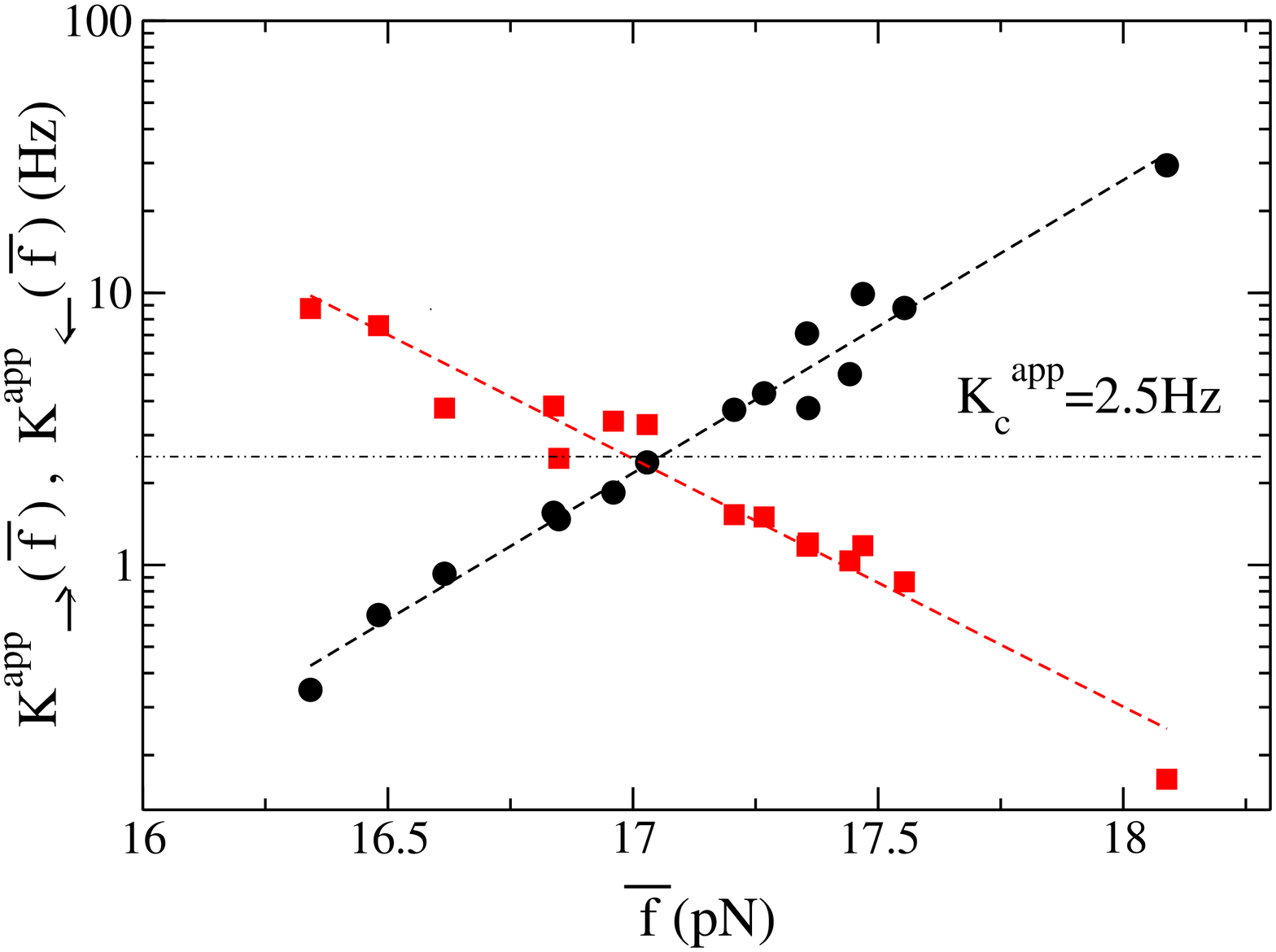}
\caption{Determination of the critical rate in the passive mode. Rates in the passive mode as a function of the
average force $\overline{f}$. The dashed lines are fits to the
simplified rates \eref{e7}. A least squares fit to the data gives $x^{\rm
  F}=10.2\ {\rm nm},x^{\rm UF}=8.7\ {\rm nm}, \mu=0.1, K\sub{c}^{\rm app}=2.5\ {\rm
  Hz}$. Data have been obtained for 3 molecules.}
\label{fig7}
\end{center}
\end{figure}

Hopping experiments are useful to directly test the validity of the
Kramers--Bell simplified kinetic rates \eref{e7}and extract kinetic
parameters such as the coexistence rate $\overline{k}\sub{c}$ and the distances
between the TS and the F and U states, $x^{\rm F}$ and $x^{\rm
  UF}$. A typical hopping trace
is shown in \fref{fig6} (left panel). In the passive mode the trap is in a
fixed position and the force switches between two values as the molecule
executes transitions between the F and the UF states
\cite{WenManLiSmiBusRitTin07,ManWenLiSmiBusTinRit07}. A typical
histogram of the forces is shown in \fref{fig6} (right panel). By fitting
the histograms to Gaussians, we can determine the average force in the F
and UF states, $f\sub{F},f\sub{UF}$, and the average force
$\overline{f}=(f\sub{F}+f\sub{UF})/2$ for each passive mode
configuration. From the dichotomic traces shown in \fref{fig6}a
we can also measure the residence times in the F and UF states for each
trap position, $\tau\sub{F}(\overline{f}),\tau\sub{UF}(\overline{f})$. By averaging
($\langle\dots\rangle$) the residence times we can extract the passive
apparent rates,
\be
K_{\rightarrow}^{\rm app}(\overline{f})=\frac{1}{\langle
  \tau\sub{F}(\overline{f})\rangle}\,,\qquad K_{\leftarrow}^{\rm app}(\overline{f})=
\frac{1}{\langle \tau\sub{UF}(\overline{f})\rangle}\,.
\label{passiverates}
\ee
The resulting rates are shown in \fref{fig7}. Linear fits to the
logarithm of the rates give the following distances: $x^{\rm F}=10.2\ 
{\rm nm},x^{\rm UF}=8.7 \ {\rm nm}$. For the coexistence rate we obtain
$K\sub{c}^{\rm app}=2.3$ Hz with $\overline{f}\sub{c}\simeq 17.1$ pN
giving $x\sub{m}=18.9$ nm and $\mu=0.1$. We also get $\Delta G=78\ k\sub{B}T$
($k\sub{B}T=4.11\ {\rm pN}\cdot {\rm nm})$. These values are compatible with
those reported from rupture force kinetic studies in
\cite{MosManForHugRit08}. 


The critical rate $\overline{k}\sub{c}$ can then be extracted from the
measured value for the passive rate $K\sub{c}^{\rm app}$ through \eref{ap12} and the rescaling factor \eref{ap10}. To evaluate $\Omega$ we need to introduce the value for $\Delta
f=f\sub{F}-f\sub{UF}\simeq 1.1$ pN and
$k\sub{B}T=4.11$ pN$\cdot$nm. Substituting these numbers into \eref{ap10} and \eref{ap12}, we
get
\be
\Omega=0.54\,, \qquad \overline{k}\sub{c}=1.4\  \mathrm{Hz}\,.
\label{omegakc}
\ee
Note that from \eref{ap16} we can also extract the value of $K\sub{c}^{\rm
  plain}$. We obtain $K\sub{c}^{\rm plain}=0.74$. As expected this agrees
with the value of the critical rate found in the representation where
rates are plotted as functions of the average force in the folded and
unfolded state (data not shown). Finally, the value for $K\sub{c}^{\rm plain}$ is not far
from the value 0.58(1) reported from rupture force kinetic studies 
(see table 1 in \cite{MosManForHugRit08}).

\subsection{Pulling experiments: results for $\langle W_{\rm dis}\rangle$ and $\langle M \rangle$}
To evaluate the average dissipated work and the average hopping number
we have carried out pulling experiments on the sequence shown in
\fref{fig1}(b) at five different pulling speeds (corresponding to
five different loading rates ranging from 1 to 16 pN/s). A summary of
the collected statistics and the results obtained for each molecule is
shown in \tref{table1}.

By measuring the mechanical work and the hopping number along the
stretching/releasing parts of each force cycle we have obtained the
average total work and the average hopping number for each set of
data. Finally, the average dissipated work for each molecule has been
measured substracting the estimated reversible work to the average
total work (how we estimated the reversible work is
explained in our companion paper \cite{MosManForHugRit08}).

In order to compare theory and experiments we take the following values
obtained from the hopping experiments previously discussed in 
\sref{hopping}: $x\sub{m}=19$ nm, $k\sub{B}T=4.11$ pN$\cdot$nm, $\mu=0.1$,
$\overline{k}\sub{c}=1.4$ Hz.

\begin{figure}
\vspace{0.9cm}
\begin{center}
\includegraphics[width=10cm,angle=0]{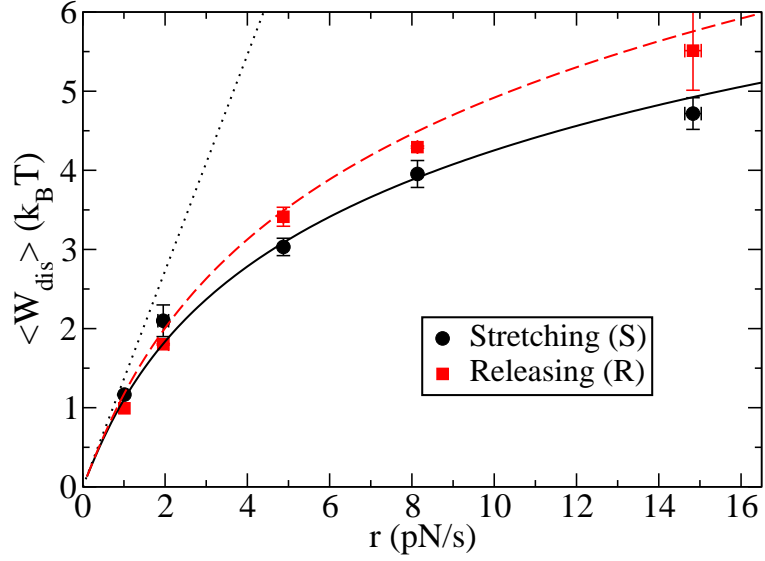}
\caption{$\langle W_{\rm dis}\rangle$  for the stretching (circles)
    and releasing process (squares) plotted as a function of the
loading rate.  The continuous (dashed) lines are the analytical
predictions \eref{en1} for the stretching (releasing) processes. The
straight dotted line is the linear response regime \eref{e9}.}
\label{fig8}
\end{center}
\end{figure}

\begin{figure}
\begin{center}
\includegraphics[width=7.5cm,angle=0]{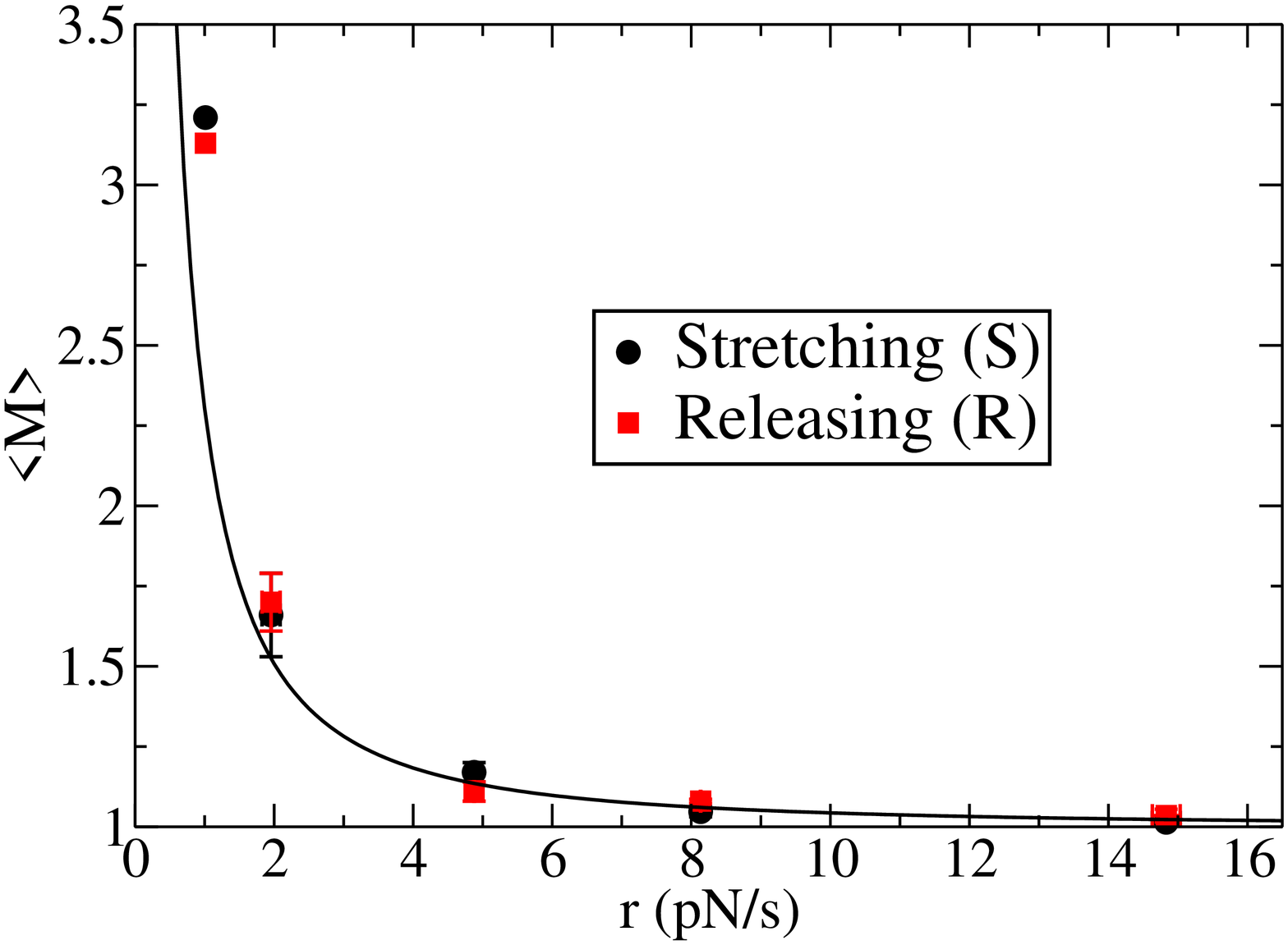}\includegraphics[width=7.5cm,angle=0]{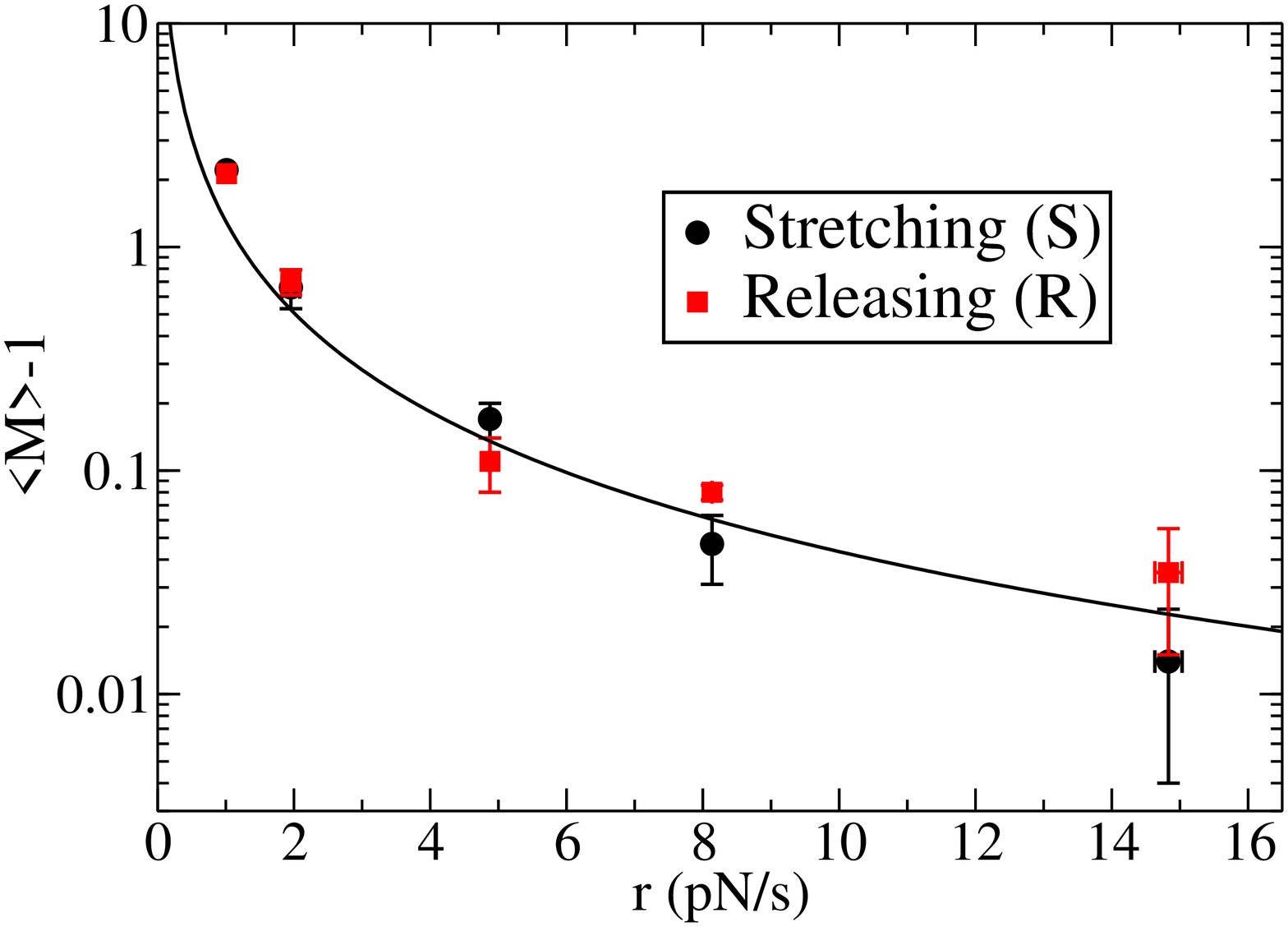}
\caption{The average hopping number $\langle M \rangle$  for the stretching (circles)
    and releasing (squares) processes as a
function of the loading rate. The continuous line is the analytical
prediction \eref{en2} for the stretching and releasing processes that
satisfy the symmetry relation \eref{sym2}. (Left panel)  $\langle M \rangle$ in normal
scale. (Right panel) To better appreciate the agreement between theory
and experiments we plot the quantity $\langle M \rangle-1$ in
logarithmic scale.}
\label{fig9}
\end{center}
\end{figure}

\begin{figure}
\begin{center}
\vspace{0.9cm}
\includegraphics[width=10cm,angle=0]{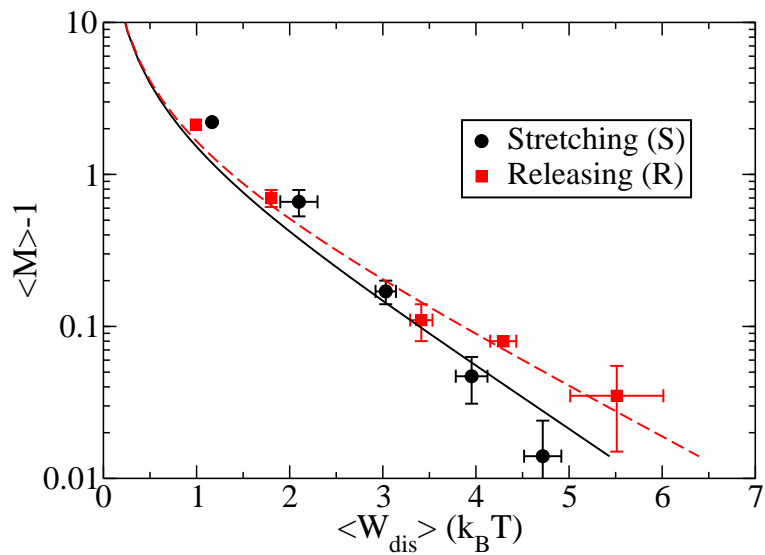}
\caption{The average hopping number $\langle M \rangle$ as a
function of $\langle W_{\rm dis}\rangle\sub{S}$ and  $\langle W_{\rm
dis}\rangle\sub{R}$ for the stretching (circles) and releasing (squares) processes.}
\label{fig10}
\end{center}
\end{figure}

\begin{figure}
\begin{center}
\includegraphics[width=10cm,angle=0]{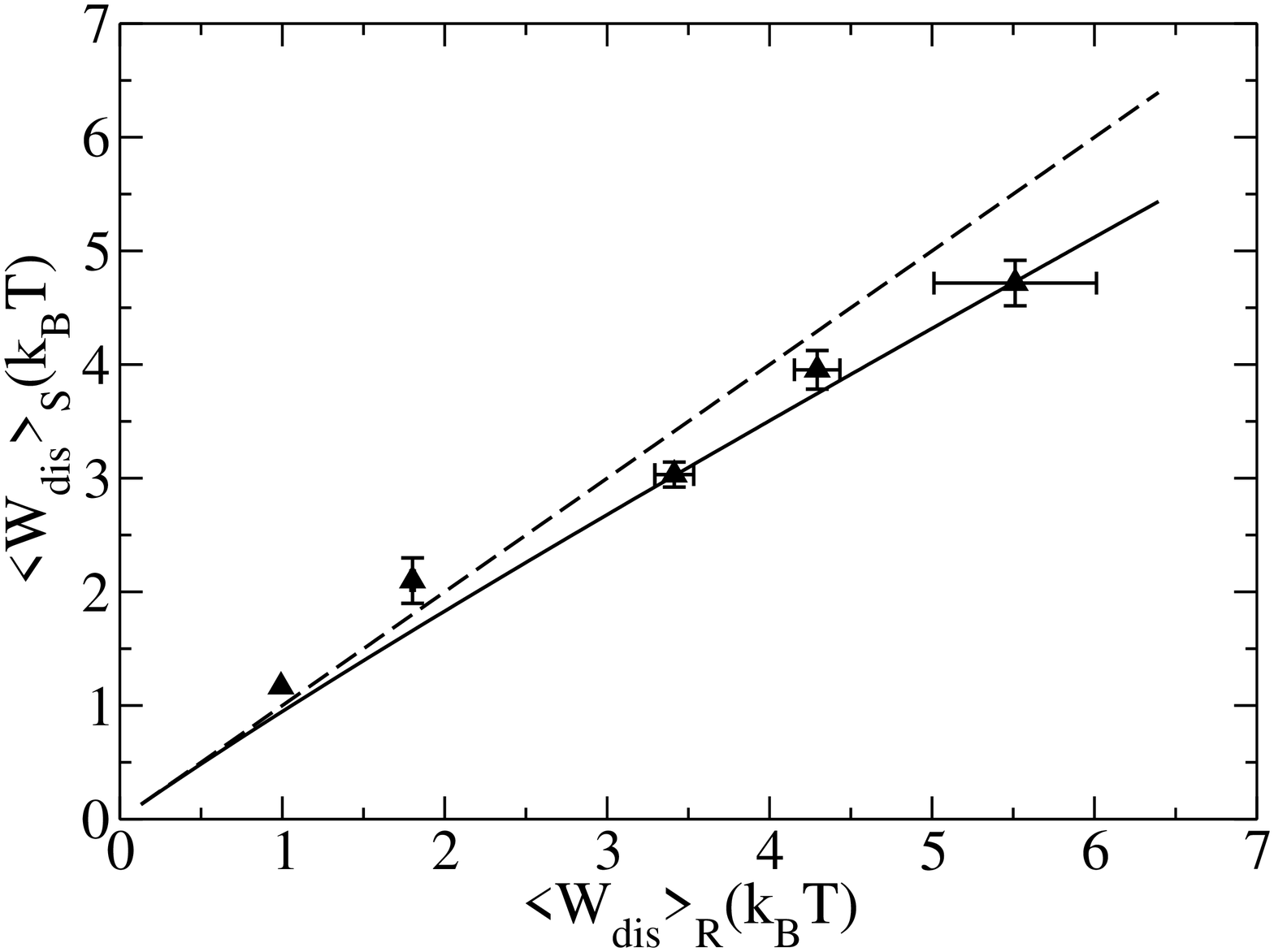}
\caption{ The average dissipated work $\langle W_{\rm dis}\rangle\sub{S}$
    as a function of $\langle W_{\rm dis}\rangle\sub{R}$. The straight line
    corresponds to the case $\mu=0$ that separates fragile and
    compliant behavior.}
\label{fig11}
\end{center}
\end{figure}

In \fref{fig8}, \fref{fig9}, \fref{fig10}, \fref{fig11} we
show various plots of the average dissipated work and the average
hopping number at five pulling speeds.The experimental results
for the dissipated work are in agreement with the theory over a wide range
of pulling speeds (\fref{fig8}). Moreover, the symmetry relation
\eref{sym2} seems reasonably well satisfied by the experimental
data. The dependence of $\langle M\rangle$ on the loading rate is well reproduced by the theory 
(\fref{fig9}). The agreement is specially good in the right panel of figure
\fref{fig9} where the asymptotic large $r$ region where $\langle M \rangle\to 1$ has been
expanded by plotting $\langle M \rangle-1$ in logarithmic scale as a function of the
loading rate $r$. 

Finally, in \fref{fig10} we plot the average dissipated work as
a function of the average hopping number whereas in \fref{fig11}
we plot the average dissipated work along the stretching process as a
function of the average dissipated work along the release
process. The plots shown in \fref{fig10} and \fref{fig11}
have in common that depend on the value of $\mu$ but not on
  $\overline{k}\sub{c}$. Therefore, just by measuring $\langle W_{\rm
  dis}\rangle$ and $\langle M \rangle$ along the stretching and
releasing processes we might be able to easily identify the value of
the fragility of the molecule.  The strong dispersion observed in the
different {\it isofragility} lines shown in \fref{fig4} (left
bottom panel) together with the unavoidable experimental errors in the
data (see \fref{fig10}), preclude the use of a plot $\langle M
\rangle$ vs.\ $\langle W_{\rm dis}\rangle\sub{S}$ to infer the value of
$\mu$. More convenient is the plot of $\langle W_{\rm dis}\rangle\sub{S}$
vs.\ $\langle W_{\rm dis}\rangle\sub{R}$ where the {\it isofragility} lines
change monotonically with $\mu$ (see \fref{fig4}, right bottom
panel) making easier to extract the
value of $\mu$ from the experimental data (\fref{fig11}).

\section{Conclusions}
We have investigated irreversibility and dissipation
effects in molecules that fold/unfold in a two state manner under the
action of mechanical force. We have developed a general theory for two
state molecules capable of predicting the average dissipated work and
the average hopping number as a function of the loading rate and other
parameters that describe the geometrical features of the free energy
landscape of the molecule. The average dissipated work and the average
hopping number are shown to depend on only two parameters: the fragility
$\mu$ (which characterizes how brittle and compliant is the molecular
structure) and the coexistence rate $\overline{k}\sub{c}$ (measured at the
coexistence force where the free energies of the folded and the unfolded
states are equal). These two parameters can be experimentally measured in hopping
experiments. We have tested the theory by carrying out pulling experiments on
DNA hairpins using optical tweezers. We remark two interesting results of the present work: 

\begin{itemize}

\item{\bf Rescaling of the kinetic rates.} We showed that in
  pulling experiments where the cantilever or the trap is moved at a
  constant pulling rate, the hopping rates measured from passive mode
  hopping experiments must be rescaled in order to account for
  the appropriate experimental conditions. The derivation of the
  rescaling factor $\Omega$ \eref{ap10} shows the influence
    of the experimental setup to quantify ireversibility
    effects in small systems. In general, an accurate knowledge of
  the kinetic rates is not necessary in single molecule studies
  (often, knowing the order of magnitude of the kinetic rate is
  enough). However, in our work it is necessary to determine the
  kinetic rates with high accuracy (let us say within 10\%) in
  order to get an adequate comparison between theory and
  experiments. The rescaling factor $\Omega$, albeit of order
  one, introduces an important correction that makes theory predictive
  in pulling experiments where the force is not controlled (e.g. in
  optical tweezers or AFM experiments).

\item{\bf Symmetry property of hopping number.} The symmetry
property for the average hopping number \eref{sym2} appears as an
interesting non-trivial consequence of the detailed
balance property. We can define the distributions
$p\sub{S(R)}(M)$ for the fraction of trajectories with $M$ transitions,
$\langle M\rangle$ being just the first moment of the distributions. Is such distribution
identical for the stretching and releasing processes, $p\sub{S}(M)=p\sub{R}(M)$?  A more elaborated
theory, in the lines of the work recently developed by Chvosta and
collaborators \cite{SubChv07}, might provide the answer to this question. Also it is
interesting to speculate whether such a result holds in general reaction
pathways beyond the two-state approximation.

\end{itemize}

\begin{figure}
\begin{center}
\includegraphics[width=10cm,angle=0]{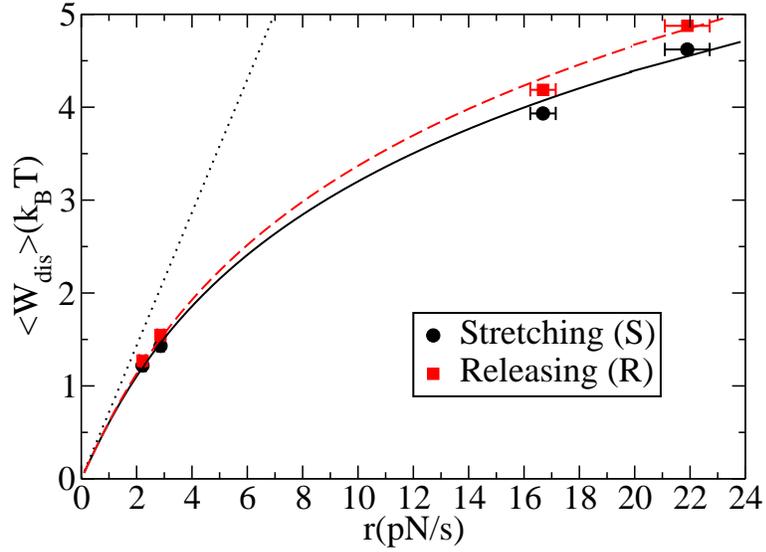}
\caption{$\langle W_{\rm dis}\rangle$ for the stretching (circles)
    and releasing process (squares) plotted as a function of the loading
    rate in a 19 base pairs DNA sequence that was investigated 
    \cite{MosForHugRit08} and with kinetics that deviates from perfect two-states
    behavior.  The continuous (dashed) lines are the analytical
    predictions \eref{en1} for the stretching (releasing) processes. The
    straight dotted line is the linear response regime \eref{e9}. The
    parameters for the simplified rates that are extracted from 
    hopping experiments in the passive mode are: $x^{\rm
      F}=9.53$ nm, $x^{\rm UF}=9.00$ nm, $\Delta f=1.0$ pN and
    $K\sub{c}^{\rm app}=4.4$ Hz. These parameters give $\mu=0.02,\Omega=0.57$ and
    $\overline{k}sub{c}=2.5$ Hz.}
\label{fig12}
\end{center}
\end{figure}

The agreement found between theorectical predictions and experiments
in DNA hairpins validates our theory and shows how this can be applied
to infer kinetic information of the molecule under study. In general,
the current theortical results should be valid for RNAs and proteins
as well. However many biomolecules do not fold into a two-state manner
so the current theory should not be applicable to systems
characterized by complex free energy landscapes such as molecules with
more than two states (e.g. molecules with intermediates or misfolded
states) or even in two-state folders with many kinetics barriers. Yet
we expect that the limits of applicability of the current theory
should be flexible enough to describe a wide range of cases reasonably
well. In \fref{fig12} we compare the theory and the experiments
for a 19 bps DNA sequence \cite{MosManForHugRit08} that is not a
faithful model of a two-state folder. The free energy landscape
calculated for this molecule shows that the F and UF states are
separated by two kinetic barriers \cite{MosForHugRit08}. Still the
theory applied to such sequence agrees pretty well with the
experiments. The extension of the current theory to more complex free
energy landscapes as well as establishing limitations of the present theory
remain as interesting open questions for future research.

\ack
We are grateful to Anna Alemany for a careful reading
of the manuscript. We acknowledge financial support from grants
FIS2007-61433, NAN2004-9348, SGR05-00688.

\appendix

\section{{Distribution of the work and the hopping number in a two-state system \label{NW_C}}} 
We want to compute the distribution of probability of the work \eref{Ba}, and the hopping number \eref{Bb}, given by \eref{Bc} and \eref{Bd}. 
We use the integral representation of the delta function,
$\delta (x)=\frac{1}{2\pi}\int_{-\infty}^{\infty}\rmd \lambda \exp(\rmi \lambda x)$, and introduce in \eref{Bc}
 the following identity:
\be
1=\prod_{k=0}^{N\sub{s}-1}\frac{1}{2\pi}\int_{-\infty}^{\infty}\rmd\gamma ^{k}\rmd m^{k}\exp \left[\rmi \gamma^{k}\left(m^{k}-\frac{1}{N}\sum_{i=1}^{N}\sigma_{i}^{k}\right)\right]\,. 
\label{Ca}
\ee
After some manipulations, the distribution of any observable $\theta$ (e.g. work or hopping number) 
can be written as:
\be
P_{N}(\theta )\propto \int_{-\infty}^{\infty}\rmd \lambda \prod_{k=0}^{N\sub{s}-1}\rmd \gamma ^{k}\rmd m^{k}\exp \left[Nf_{\theta}\left(w,\lambda, \{ \gamma ^{k}\} ,\{ m^{k}\} \right)\right] \,,
\label{Cb}
\ee
where $f_{\theta}$ is a function that depends on the observable $\theta$. We will use the notation 
$f\sub{M}=a$ and $f\sub{W}=b$, for the work and the hopping number respectively. 
In the continuous limit the probability distribution \eref{Cb}, becomes a path integral, 
where the $a$ and $b$ functions read:
\be
\fl a=&-\lambda \left\{W-x\sub{m}/2\left[\left(f\sub{max}-f\sub{min}\right)+\int_{0}^{t}\rmd s\,m(s)r(s)\right]\right\}+ \nonumber \\
\fl &+\frac{1}{2}\int_{0}^{t}\rmd s\left\{m(s)\left[2\dot{\gamma}(s)+c(s)\right]+d(s)\right\}+ 
\log \left[\rme^{\gamma^{0}}k_{\rightarrow}(f\sub{min})+\rme^{-\gamma^{0}}k_{\leftarrow}(f\sub{min})\right]\,,
\label{Cc} \\
\fl b=&-\lambda \frac{N}{2}+\frac{1}{2}\int_{0}^{t}\rmd s\left\{m(s)\left[2\dot{\gamma}(s)+e(s)\right]+g(s)\right\}+ \log\left[\rme^{\gamma^{0}}k_{\rightarrow}(f\sub{min})+
\rme^{-\gamma^{0}}k_{\leftarrow}(f\sub{min})\right] \label{Ce}
\ee
The function $r(s)$ is the rate of increasing the external field $f$, $r(s)=\frac{\rmd f}{\rmd s}$,
 and $f\sub{min}$ is the initial value of the field in the ramping
protocol. $k_{\rightarrow}$ and $k_{\leftarrow}$ are the rates \eref{e7}
corresponding to the transition from $\sigma=-1$ to $\sigma=1$ and from $\sigma=1$ to 
$\sigma=-1$, respectively. 
The functions $c(s)$, $d(s)$,
$e(s)$ and $g(s)$ are given by:
\be
c(s)&=k_{\leftarrow}(f(s)) (\rme^{-2\gamma (s)}-1)-k_{\rightarrow}(f(s))(\rme^{2\gamma (s)}-1)\,,\label{Cf1}\\   
d(s)&=k_{\leftarrow}(f(s)) (\rme^{-2\gamma (s)}-1)+k_{\rightarrow}(f(s))(\rme^{2\gamma (s)}-1)\,,\label{Cf2}\\
e(s)&=k_{\leftarrow}(f(s))(\rme^{-2\gamma (s)+\lambda /2}-1)-k_{\rightarrow}(f(s))(\rme^{2\gamma (s)+\lambda /2}-1)\,,\label{Cg1}\\   
d(s)&=k_{\leftarrow}(f(s))(\rme^{-2\gamma (s)+\lambda /2}-1)+k_{\rightarrow}(f(s))(\rme^{2\gamma (s)+\lambda /2}-1)\,,\label{Cg2}
\ee
with boundary conditions $\gamma (t)=0$ and $m(0)=\tanh(\gamma(0)+\beta x\sub{m}f\sub{min}/2)$. 
Note that these boundary conditions break causality.
In the large $N$ approximation, the integral given by \eref{Cb} can be estimated 
by using the saddle point technique. The saddle point equations, obtained by maximizing the $a$ and $b$ 
functions over the variables $\lambda$, $\gamma(s)$ and $m(s)$, are given by:
\be
\frac{\partial a}{\partial \lambda}&=0=W-x\sub{m}/2\left[(f\sub{max}-f\sub{min})-\int_{0}^{t}\rmd s\, m(s)r(s)\right]\,,\nonumber\\
\frac{\delta a}{\delta \gamma(s)}&=0=\dot{m}(s)+m(s)\left[d(s)+k_{\rm T}(f(s))\right]+c(s)+k_{\rm M}(f(s))\,,\nonumber\\
\frac{\delta a}{\delta m(s)}&=0=-\frac{\gamma (s)x\sub{m}r(s)}{2}+\dot{\gamma}(s)+\frac{c(s)}{2}\,,\nonumber\\
\frac{\partial b}{\partial \lambda}&=0=\frac{N}{2}+\frac{1}{4}\int_{0}^{t}\rmd s\left\{m(s)\left[e(s)+
k_{\rm M}(f(s))\right]+g(s)+k_{\rm T}(f(s))\right\}\,,\nonumber\\
\frac{\delta b}{\delta \gamma(s)}&=0=\dot{m}(s)+m(s)\left[g(s)+k_{\rm T}(f(s))\right]+e(s)+k_{\rm M}(f(s)))\,,\nonumber\\
\frac{\delta b}{\delta m(s)}&=0=\dot{\gamma}(s)+\frac{e(s)}{2}\,,\nonumber
\ee
where the dots mean the derivative with respect to the time $s$, and $k_{\rm T}(f),k_{\rm M}(f)$ are given by
 $k_{\rm T}(f)=k_{\rightarrow}(f)+k_{\leftarrow}(f)$, $k_{\rm M}(f)=k_{\leftarrow}(f)-k_{\rightarrow}(f)$. 
It can be shown \cite{Rit22} that the 
most probable value for the observable $\theta$, $\theta^{+}$, 
can be obtained as $\frac{\partial f_{\theta}}{\partial\theta}{\mid}_{\theta^{+}}=0$, which gives $\lambda^{+}=0$.
In the large $N$ limit the most probable value and the average value coincide.
The average work and hopping number read: 
\be
\langle W\rangle =(x\sub{m}/2)\left[(f\sub{max}-f\sub{min})-\int_{f\sub{min}}^{f\sub{max}}m(f)\rmd f\right]\,,
\label{Ce30} \\
\langle M\rangle =(1/2)\int_{f\sub{min}}^{f\sub{max}}\left[m(f)k_{\rm M}(f)+k_{\rm T}(f))(1/r(f)\right]\rmd f\,,
\label{Ce6}
\ee
where the functions $m\sub{eq}(f)$ and $m(f)$ are given by:
\be
m\sub{eq}(f)=\frac{k_{\rm M}(f)}{k_{\rm T}(f)}\,, 
\label{Ce4}
\ee
and
\be
m(f)=m\sub{eq}(f)-\int_{f\sub{min}}^{f}\frac{\rmd m\sub{eq}(f_{1})}{\rmd f_{1}}\exp\left[-\int_{f_{1}}^{f}\frac{k_{\rm T}(f_{2})}{r(f_{2})}\rmd f_{2}\right]\rmd f_{1} \,.
\label{Ce5}
\ee

\section{Expansion of the work and the hopping number in the low loading rate regime \label{NW_D}} 

For a system described by the folding and unfolding rates, 
$k_{\leftarrow}(f)$ and $k_{\rightarrow}(f)$,  given in \eref{e7}, 
the functions $k_{\rm T}(f)$ and $k_{\rm M}(f)$ read:
\be
k_{\rm T}(f)=k_{\leftarrow}(f)+k_{\rightarrow}(f)=2k\sub{c}\rme^{-\beta \mu x}\cosh x\,,
\ee
and
\be
k_{M}(f)=k_{\leftarrow}(f)-k_{\rightarrow}(f)=2k\sub{c}\rme^{-\beta \mu x}\sinh x \,,
\ee
where $k\sub{c}$ is the folding or unfolding rate at the coexistence force $F^{rm c}$, $x$ corresponds to a dimensionless force $x=\frac{(f-F^{\rm c})x\sub{m}}{2k\sub{B}T}$, and $\mu$ is defined as $\mu=\frac{x^{\rm F}-x^{\rm UF}}{x\sub{m}}$. 
Using these expressions we can write 
the average dissipated work and the average hopping number, \eref{e3} and \eref{e6}, as: 
\be
\langle W_{\rm dis}\rangle =k\sub{B}T \int_{-\infty}^{\infty}\rmd x\, F(-\infty,x)\,, 
\ee
with
\be
F(-\infty,x)=&\int_{-\infty}^{x}\rmd y 
\chi\sub{eq}(y)\rme^{-\frac{\phi (y,x)}{\tilde{r}}}\,,
\label{QD1} \\
\langle M\rangle =&\frac{1}{2 \tilde{r}}\left[\int_{-\infty}^{\infty}\rmd x(-m\sub{eq}^{2}(x)+1)\rme^{-\beta \mu x}\cosh x \right. \nonumber \\ 
&\left.+\int_{-\infty}^{\infty}\rmd x(\rme^{-\beta \mu x}\sinh x)F(-\infty,x)\right]\,,
\label{QD2}
\ee
where $\tilde{r}$ is a dimensionless loading rate, given by $\tilde{r}=\frac{x\sub{m}r}{k\sub{c}4k\sub{B}T}$, 
 the function $m\sub{eq}$ (defined in \eref{e4}) is given by $m\sub{eq}(x)=\tanh x$,  
and $\chi\sub{eq}(x)=\frac{\rmd m\sub{eq}(x)}{\rmd x}=\frac{1}{\cosh^{2}x}$.
Finally, the function $\phi$ is given by $\phi(y,x)=-\int_{y}^{x}\rmd z \frac{k_{\rm T}(z)}{2k\sub{c}}=\int_{y}^{x}\rmd z \,\rme^{-\mu z}\cosh z$.
Since $\phi(y,x)$ is defined positive for all $x\geq y$, when $r$ goes to zero (or $\tilde{r}\rightarrow 0$) 
 $\rme^{-\phi(y,x)/\tilde{r}}$ vanishes except in $x=y$ where $\phi=0$. Therefore, at the low 
loading rate regime, we can use a saddle point approximation and expand the integrand of $F(-\infty,x)$:
\be
\fl F(-\infty,x)&=&\int_{-\infty}^{x}\rmd y[\chi\sub{eq}(x)-\chi'\sub{eq}(x)(y-x)+\frac{\chi''\sub{eq}(x)(y-x)^{2}}{2}+\dots]\nonumber\\
\fl & &\cdot \rme^{\frac{-\phi'(x) (y-x)}{\tilde{r}}} 
[1+\frac{\phi''(x)}{2\tilde{r}}(x-y)^2-\frac{\phi'''(x)}{6\tilde{r}}(y-x)^3+\dots]\nonumber\\
\fl&=&\tilde{r} \frac{\chi\sub{eq}(x)}{\phi'(x)}+
\tilde{r}^{2}\left[\frac{\phi''(x)\chi\sub{eq}(x)}{\phi '(x)^{3}}-\frac{\chi'\sub{eq}(f)}{\phi '(x)^{2}}\right] \nonumber \\
& &+\tilde{r}^{3}\left[-\frac{3\chi\sub{eq}'(f)\phi ''(x)}{\phi '(x)^{4}}+\frac{\chi ''\sub{eq}(f)}{4\phi '(x)}\right]+\Or(\tilde{r}^{4}) \,,
\ee
where the prime means the derivative with respect to $x$.
By introducing the previous expansion to \eref{QD1} and \eref{QD2} we get:
\begin{eqnarray}
\beta \langle W_{\rm dis}\rangle =A (\mu) \tilde{r}+B (\mu) \tilde{r}^{2}+C(\mu)\tilde{r}^{3}+O(\tilde{r}^{4})~,\nonumber\\
\langle M\rangle =\frac{\alpha (\mu)}{\tilde{r}}+\beta (\mu) +\gamma (\mu)\tilde{r}+\delta (\mu)\tilde{r}^{2}+ O(\tilde{r}^{3})~.
\end{eqnarray}
The different coefficients of the expansion are given by: 
\begin{eqnarray}
A(\mu)=-\int_{-\infty}^{\infty}dx \frac{e^{\mu x}}{{\rm cosh}^{3}x}=\frac{\pi}{2}(1-\mu^{2})\sec (\pi\mu/2)~,
\end{eqnarray}
\begin{eqnarray}
B(\mu)=\int_{-\infty}^{\infty}dx e^{-2\mu x}{\rm sech}^{4}y(\mu +3{\rm tanh}x)=\frac{-2}{3}\mu^{2}(1-\mu^{2})\pi\csc(\pi\mu)~,
\end{eqnarray}
\begin{eqnarray}
C(\mu)&=&\int_{-\infty}^{\infty}dx \frac{e^{-3\mu x}}{{\rm cosh}^{5}x}(\frac{-3}{{\rm cosh}^{2}x}+9{\rm tanh}^{2}x+4\mu{\rm tanh}x-\mu^{2})\nonumber\\
&=&\frac{3}{40}(-5+51\mu^{2}-55\mu^{4}+ 9\mu^{6})\pi\sec(3\pi\mu/2)~,
\end{eqnarray}
\begin{eqnarray}
\alpha (\mu)=\frac{1}{2}\int_{-\infty}^{\infty}dx \frac{e^{\mu x}}{{\rm cosh}x}=\frac{\pi}{2}\sec (\pi\mu/2)~,
\end{eqnarray}
\begin{eqnarray}
\beta (\mu)=0~,
\end{eqnarray}
\begin{eqnarray}
\gamma (\mu)&=&1/2\int_{-\infty}^{\infty}dx e^{-\mu x}{\rm sech}^{3}x {\rm tanh}x (\mu+3{\rm tanh}x)\nonumber\\
&=&\frac{1}{48}(9-10\mu^{2}+\mu^{4})\pi\sec(\pi\mu/2)~,
\end{eqnarray}
\begin{eqnarray}
\delta (\mu)&=& \frac{1}{2} \int_{-\infty}^{\infty}dx \frac{{\rm sinh}x e^{-2\mu x}}{{\rm cosh}^{5}x} (\frac{-3}{{\rm cosh}^{2}x}+\nonumber\\
&+&9{\rm tanh}^{2}x+4\mu{\rm tanh}x-\mu^{2}) =-1/3(\mu-\mu^{3})\pi\csc(\pi\mu).
\end{eqnarray}

\end{document}